\documentclass[pra,aps]{revtex4}
\newcommand{\be}{\begin{equation}}
\newcommand{\ee}{\end{equation}}
\newcommand{\brr}{\begin{eqnarray}}
\newcommand{\err}{\end{eqnarray}}
\newcommand{\nn}{\nonumber}
\newcommand{\bd}{\begin{displaymath}}
\newcommand{\ed}{\end{displaymath}}

\newcommand{\bib}{\bibitem}
\newcommand{\bfig}{\begin{figure}}
\newcommand{\efig}{\end{figure}}
\newcommand{\ie}{i.e.}

\def\alf{\alpha}
\def\bet{\beta}
\def\gam{\gamma}

\def\lam{\lambda}

\def\rpar{\right)}
\def\lpar{\left(}
\def\rbk{\right]}
\def\lbk{\left[}
\def\rbr{\right\}}
\def\lbr{\left\{}
\def\lb{\label}
\def\ro{\mbox{\boldmath $\rho$}}
\def\sig{\mbox{\boldmath $\sigma$}}
\def\sqq{\mbox{\boldmath $\mathfrak{X}$}}
\def\sop{\mbox{\boldmath $\mathcal{S}$}}
\def\rg{\rangle}
\def\lgg{\langle}
\def\nc{\mathrm{i}}
\def\coloneq{\mathrel{\mathop:}=}

\def\fa{\mathfrak{a}}
\def\fz{\mathfrak{z}}
\def\fs{\mathfrak{s}}
\def\ns{\mbox{\scriptsize $N$}}
\def\half{\frac{1}{2}}
\usepackage{ifvtex}
\usepackage{ifpdf}
\usepackage[dvips]{graphicx}
\usepackage{amssymb}
\usepackage{mathrsfs}
\usepackage{upmath}
\usepackage{yfonts}
\usepackage{dsfont}
\begin{document}
%
\title{Discrete squeezed states for finite-dimensional spaces}
\author{Marcelo A. Marchiolli, Maurizio Ruzzi, Di\'{o}genes Galetti}
\affiliation{Instituto de F\'{\i}sica Te\'{o}rica, Universidade Estadual Paulista, \\
             Rua Pamplona 145, 01405-900, S\~{a}o Paulo, SP, Brazil \\
             E-mail address: mamarchi@ift.unesp.br, mruzzi@ift.unesp.br, galetti@ift.unesp.br}
\date{\today}
%
\begin{abstract}
\vspace*{0.1mm}
\begin{center}
\rule[0.1in]{142mm}{0.4mm}
\end{center}
We show how discrete squeezed states in an $N^{2}$-dimensional phase space can be properly constructed out of the finite-dimensional
context. Such discrete extensions are then applied to the framework of quantum tomography and quantum information theory with the aim
of establishing an initial study on the interference effects between discrete variables in a finite phase-space. Moreover, the
interpretation of the squeezing effects is seen to be direct in the present approach, and has some potential applications in different
branches of physics.
\vspace*{0.1mm}
\begin{center}
\rule[0.1in]{142mm}{0.4mm}
\end{center}
\end{abstract}
\maketitle

\section{Introduction}

The first prototypes of continuous wavepackets related to the coherent and squeezed states appeared in the literature in 1926 through
the publication of three seminal papers by Schr\"{o}dinger \cite{r1}, Kennard \cite{r2}, and Darwin \cite{r3}, where the evolutions of
Gaussian wavepackets for an oscillator, a free particle and a particle moving in uniform constant electric and magnetic fields were
treated for the first time. In fact, they represent not only an initial historic mark in the study of nonclassical states of the
electromagnetic field, but also, from the contemporary point of view, the cornerstones of modern quantum optics \cite{r4}. Since then a
huge number of works dedicated to application, generation, and detection of nonclassical states have been published in many specialized
scientific journals covering different areas of knowledge in physics (such as, for instance, solid state physics, nuclear physics, high
energy physics, general relativity, and cosmology). Nowadays, beyond these fundamental features, these states have also potential
applications in quantum information theory and quantum computation since continuous-variable entanglement can be efficiently produced
using squeezed light and nonlinear optics \cite{r5,r6}. However, if one considers physical systems with a finite-dimensional space of
states, a sound theoretical framework must be employed in order to describe properly the nonclassical states. In this sense, it is worth
noticing that there are various formalisms proposed in the literature for finite-dimensional Hilbert spaces with convenient inherent
mathematical properties which can be applied in the description of these quantum systems
\cite{r7,r8,r9,r10,r11,r12,r13,r14,r15,r16,r17,r18,r19}. In particular, let us focus our attention upon the approach developed in Ref.
\cite{r11} for the discrete representatives of the coherent states which has its algebraic structure based on the technique of
constructing unitary operator bases initially formulated by Schwinger \cite{r20}. Basically, this approach consists in proposing a new
specific operator basis (whose discrete labels are congruent modulo $N$) which allows the direct mapping of Weyl-Wigner representatives
in an $N^{2}$-dimensional phase space through a simple trace operation; in addition, a generator of discrete displacements for this
particular phase space, which exhibits an additional topological phase related to the mod$(N)$ invariance, is also derived and its
properties properly discussed in details. Subsequently, we presented in Ref. \cite{r18} an {\it ab initio} construction that inherently
embodies the discrete analogues of the desired properties of the Cahill-Glauber formalism \cite{r21}, where some direct applications in
quantum information processing, quantum tomography, and quantum teleportation were explored. Furthermore, this specific construction
process opens the possibility of dealing with discrete wavepackets whose widths are modified, in principle, by parameters that mimic
the continuous squeezing effect.

The main goal of this paper is to fill this breach by constructing a consistent formalism for the discrete squeezed states in
finite-dimensional spaces. For this purpose, we first present briefly some useful results obtained in Ref. \cite{r18} for
finite-dimensional Hilbert spaces with special emphasis on the mod$(N)$-invariant operator basis and the discrete counterpart of the
nondiagonal projector related to the coherent states. Next, we define a wavepacket for the discrete squeezed-vacuum state in connection
with the equivalence relation between the wavepacket phase space representation and the phase space generated by the continuous
squeezed states \cite{r22}. Consequently, this definition leads us to study two special classes of discrete squeezed states in an
$N^{2}$-dimensional phase space and establish some results for their respective diagonal projectors in complete analogy to the
continuous ones. Such discrete extensions are then applied to the context of quantum tomography and quantum information theory with the
aim of enlarging our knowledge on the quantum interference effects in a discrete phase space. In particular, we attain new results
within which some of them deserve to be mentioned: (i) we show that a generalized version of the scattering circuit introduced in Ref.
\cite{r23} can be used to measure any discrete Wigner function and/or its corresponding characteristic function in the presence of
squeezing effects; (ii) we establish a first reasonable measure of functional correlation between the discrete variables of an
$N^{2}$-dimensional phase space which can be extended within the context of quantum information theory, as a measure of entanglement in
multipartite systems \cite{r24}, or in other branches of physics whose systems of interest can be described by finite state spaces; and
finally, (iii) we apply this measure for the discrete squeezed-vacuum state and verify that its maximum value depends essentially on
the Hilbert-space dimension.

This paper is organized as follows. In Section II we discuss briefly the extension of the Cahill-Glauber formalism for
finite-dimensional Hilbert spaces with emphasis on some essential features exhibited by the mod$(N)$-invariant operator basis. In
Section III we introduce the discrete representatives of the displaced squeezed-vacuum states and squeezed coherent states in a finite
phase-space, where their respective diagonal projectors are derived in details. The results obtained are then applied in Section IV
within the context of quantum tomography and quantum information theory in order to establish two different but complementary methods
for studying quantum interference effects on such phase space. Finally, Section V contains our summary and conclusions. Furthermore, we
add two mathematical appendices related to the calculational details of some basic expressions used in the previous sections. 

\section{Preliminaries}

The cornerstone of the extended Cahill-Glauber formalism for finite-dimensional spaces is given by the mod($N$)-invariant operator basis
\cite{r18}
\be
\lb{e1}
{\bf T}^{(s)}(\mu,\nu) \coloneq \frac{1}{\sqrt{N}} \sum_{\eta,\xi = - \ell}^{\ell} \exp \lbk - \frac{2 \pi \nc}{N} (\eta \mu + \xi \nu)
\rbk {\bf S}^{(s)}(\eta,\xi) ,
\ee
which consists of a discrete double Fourier transform of the extended mapping kernel ${\bf S}^{(s)}(\eta,\xi) = [ \mathscr{K}(\eta,\xi) 
]^{-s} {\bf S}(\eta,\xi)$, where the labels $\eta$ and $\xi$ represent the dual momentum and coordinate variables of a discrete
$N^{2}$-dimensional phase space. In particular, these labels are congruent modulo $N$ and assume integer values in the symmetrical
interval $[-\ell,\ell]$ for $\ell = (N-1)/2$. The extra term $\mathscr{K}(\eta,\xi)$ is defined in this context through the ratio
$\mathscr{M}(\eta,\xi) / \mathscr{M}(0,0)$,
\brr
\mathscr{M}(\eta,\xi) &=& \frac{\sqrt{\fa}}{2} \Bigl[ \vartheta_{3} (\fa \eta | \nc \fa) \vartheta_{3} (\fa \xi | \nc \fa) +
\vartheta_{3}(\fa \eta | \nc \fa) \vartheta_{4} (\fa \xi | \nc \fa) \mathrm{e}^{\nc \pi \eta} \nn \\
& & + \, \vartheta_{4} (\fa \eta | \nc \fa) \vartheta_{3} (\fa \xi | \nc \fa) \mathrm{e}^{\nc \pi \xi} + \vartheta_{4}(\fa \eta | \nc
\fa) \vartheta_{4} (\fa \xi | \nc \fa) \mathrm{e}^{\nc \pi (\eta + \xi + N)} \Bigr] \nn
\err
being the function responsible for the sum of products of Jacobi theta functions evaluated at integer arguments \cite{r25}, with 
$\fa = (2N)^{-1}$ fixed. Furthermore, the complex parameter $s$ obeys $| s | \leq 1$, and 
\bd
{\bf S}(\eta,\xi) \coloneq \frac{1}{\sqrt{N}} \exp \lpar \frac{\nc \pi}{N} \eta \xi \rpar {\bf U}^{\eta} {\bf V}^{\xi} 
\ed
corresponds to the symmetrized version of the unitary operator basis proposed by Schwinger \cite{r20}. Finally, it is worth mentioning
that a comprehensive and useful compilation of results and properties of the unitary operators ${\bf U}$ and ${\bf V}$ can be promptly
found in Ref. \cite{r11}; here the main focus of our attention is related to the essential features exhibited by Eq. (\ref{e1}).

For instance, the set of $N^{2}$ operators $\{ {\bf T}^{(s)}(\mu,\nu) \}_{\mu,\nu = -\ell, \ldots, \ell}$ allows us to decompose any
linear operator ${\bf O}$ through the expansion
\be
\lb{e2}
{\bf O} = \frac{1}{N} \sum_{\mu,\nu = -\ell}^{\ell} \mathcal{O}^{(-s)}(\mu,\nu) {\bf T}^{(s)}(\mu,\nu) ,
\ee
where the coefficients $\mathcal{O}^{(-s)}(\mu,\nu) \coloneq \mathrm{Tr} [ {\bf T}^{(-s)}(\mu,\nu) {\bf O} ]$ are associated with a
one-to-one mapping between operators and functions belonging to a finite phase-space characterized by the discrete labels $\mu$ and
$\nu$. The first practical application of this decomposition is related to the calculation of the mean value 
\be
\lb{e3}
\lgg {\bf O} \rg \coloneq \mathrm{Tr} ( {\bf O} \ro ) = \frac{1}{N} \sum_{\mu,\nu = - \ell}^{\ell} \mathcal{O}^{(-s)}(\mu,\nu)
F^{(s)}(\mu,\nu) ,
\ee
which leads us to define the parametrized function $F^{(s)}(\mu,\nu) \coloneq \mathrm{Tr} [ {\bf T}^{(s)}(\mu,\nu) \ro ]$ as being the
double Fourier transform of the discrete $s$-ordered characteristic function $\Xi^{(s)}(\eta,\xi) \coloneq \mathrm{Tr} [ {\bf S}^{(s)}
(\eta,\xi) \ro ]$, namely
\be
\lb{e4}
F^{(s)}(\mu,\nu) = \frac{1}{\sqrt{N}} \sum_{\eta,\xi = - \ell}^{\ell} \exp \lbk - \frac{2 \pi \nc}{N} ( \eta \mu + \xi \nu ) \rbk
\Xi^{(s)}(\eta,\xi) .
\ee
Thus, in complete analogy to the continuous case, the discrete Husimi, Wigner, and Glauber-Sudarshan functions can be directly obtained
from Eq. (\ref{e4}) for specific values of the complex parameter $s$. For physical applications associated with quantum tomography and
quantum teleportation, see Ref. \cite{r18}.

After this condensed review, we will obtain a decomposition for the nondiagonal projector related to the set of discrete coherent states
$\{ | \mu,\nu \rg \}_{\mu,\nu = - \ell, \ldots, \ell}$, whose mathematical properties were established in Ref. \cite{r11}. For this
task, let us initially consider the expansion formula for ${\bf O} = | \mu,\nu \rg \lgg \mu^{\prime},\nu^{\prime} |$ and $s=-1$ fixed.
The next step consists in substituting the mod($N$)-invariant operator basis ${\bf T}^{(-1)}(\bar{\mu},\bar{\nu})$ into Eq. (\ref{e2})
and evaluating the sums over the discrete indices $\bar{\mu}$ and $\bar{\nu}$. After lengthy calculations, the analytical expression for
the nondiagonal projector assumes the exact form
\brr
\lb{e5}
| \mu,\nu \rg \lgg \mu^{\prime},\nu^{\prime} | &=& \frac{1}{\sqrt{N}} \sum_{\eta,\xi = - \ell}^{\ell} \exp \lbr - \frac{\nc \pi}{N} \lbk
\eta (\mu + \mu^{\prime}) + \xi (\nu + \nu^{\prime}) + \mu \nu^{\prime} - \mu^{\prime} \nu \rbk \rbr \nn \\
& & \times \, \mathscr{K}(\eta - \nu + \nu^{\prime}, \xi + \mu - \mu^{\prime}) {\bf S}(\eta,\xi) ,
\err
being reduced to the diagonal projector ${\bf T}^{(-1)}(\mu,\nu) = | \mu,\nu \rg \lgg \mu,\nu |$ when $\mu^{\prime} = \mu$ and
$\nu^{\prime} = \nu$. Consequently, the probability of finding the discrete coherent state $| \mu,\nu \rg$ in the $n$th excited state
-- described in \cite{r26,r27} by a complete set of number states $\{ | n \rg \}_{n=0,\ldots,N-1}$ -- can be promptly estimated as follows:
\be
\lb{e6}
P_{n}(\mu,\nu) = \frac{1}{N} \sum_{\eta,\xi = - \ell}^{\ell} \exp \lbk - \frac{2 \pi \nc}{N} (\eta \mu + \xi \nu) \rbk \mathscr{K}
(\eta,\xi) \mathscr{K}_{n}(\eta,\xi) ,
\ee
where $\mathscr{K}_{n}(\eta,\xi) \coloneq \mathscr{M}_{n}(\eta,\xi) / \mathscr{M}_{n}(0,0)$ is connected with the diagonal matrix
element $\lgg n | {\bf S}(\eta,\xi) | n \rg$ through the equality $\mathscr{K}_{n}(\eta,\xi) = \sqrt{N} \, \lgg n | {\bf S}(\eta,\xi) |
n \rg$, and
\bd
\mathscr{M}_{n}(\eta,\xi) \coloneq \sqrt{\fa} \, \left. \frac{\upartial^{n}}{\upartial \fz^{n}} ( 1 + 2 \fz )^{-1} \mathrm{M}
(\eta,\xi;\fz) \right|_{\fz = 0} . 
\ed
In this definition, $\mathscr{M}_{n}(\eta,\xi)$ depends explicitly on the $n$-th order derivative of the product $( 1+2\fz )^{-1}
\mathrm{M}(\eta,\xi;\fz)$ evaluated at the point $\fz = 0$,
\brr
\mathrm{M}(\eta,\xi;\fz) &=& \half \Bigl\{ \vartheta_{3} [ \fa \eta | f(\fz) ] \vartheta_{3} [ \fa \xi | f(\fz) ] + \vartheta_{3} 
[ \fa \eta | f(\fz) ] \vartheta_{4} [ \fa \xi | f(\fz) ] \mathrm{e}^{\nc \pi \eta} \nn \\
& & + \, \vartheta_{4} [ \fa \eta | f(\fz) ] \vartheta_{3} [ \fa \xi | f(\fz) ] \mathrm{e}^{\nc \pi \xi} + \vartheta_{4} [ \fa \eta
| f(\fz) ] \vartheta_{4} [ \fa \xi | f(\fz) ] \mathrm{e}^{\nc \pi (\eta + \xi + N)} \Bigr\} \nn
\err
being an auxiliary function whose analytical expression differs essentially from $\mathscr{M}(\eta,\xi)$ by the presence of $f(\fz) =
\nc \fa (1-2\fz)(1+2\fz)^{-1}$ in the second argument of the Jacobi theta functions (see appendix A for technical details). Note that
for $n=0$ (reference state) the function $\mathscr{K}_{0}(\eta,\xi)$ coincides exactly with $\mathscr{K}(\eta,\xi) \coloneq \lgg \eta ,
\xi | 0,0 \rg$, as it should be in this case. Furthermore, adopting the mathematical procedure established in Ref. \cite{r28} for the
continuum limit, we also verify that $P_{n}(\mu,\nu)$ goes to 
\bd
P_{n}(q,p) = \frac{1}{n!} \exp \lpar - \frac{q^{2} + p^{2}}{2} \rpar \lpar \frac{q^{2} + p^{2}}{2} \rpar^{n}
\ed
in the limit $N \rightarrow \infty$. In the next section, we will focus upon the following question: `How can we propose a family of
squeezed states in a finite-dimensional Hilbert space by means of a discrete wavefunction whose width is modified through a squeezing
parameter?'

\section{Discrete squeezed states for finite-dimensional spaces}

The reference (vacuum) state $| 0 \rg$ has an important role within the theory of continuous coherent states \cite{r29} since they
can be generated by means of a unitary displacement operator ${\bf D}(q,p)$ acting on this reference state, namely $| q,p \rg \coloneq
{\bf D}(q,p) | 0 \rg$. Pursuing this line, we have shown in Ref. \cite{r11} that the Schwinger operator basis elements $\{ {\bf S}
(\eta,\xi) \}_{\eta,\xi = -\ell, \ldots, \ell}$ also act as displacement operators on a particular reference state to generate discrete
coherent states, according to the definition $| \mu,\nu \rg \coloneq \sqrt{N} {\bf S}(\nu,-\mu) | 0 \rg$. In this formalism, the
normalized discrete wavefunction associated with the reference state in the coordinate-like representation $\{ | u_{\gam} \rg 
\}_{\gam = -\ell,\ldots,\ell}$ is expressed as
\be
\lb{e7}
\lgg u_{\gam} | 0 \rg = \lbk \frac{2 \fa}{\mathscr{M}(0,0)} \rbk^{1/2} \vartheta_{3} (2 \fa \gam | 2 \nc \fa ) ,
\ee
where the width of the Jacobi $\vartheta_{3}$-function has a constant value equal to $\sqrt{2 \fa}$. Now, let us suppose that there
exists an analogous wavepacket for the discrete squeezed-vacuum state $| 0;\fs \rg$ whose formal expression reads as \cite{r30}
\be
\lb{e8}
\lgg u_{\gam} | 0;\fs \rg = \lbk \frac{2 \fa}{\mathscr{M}_{\fs}(0,0)} \rbk^{1/2} \vartheta_{3} (2 \fa \gam | 2 \nc \fa \fs^{-2}) ,
\ee
with $\mathscr{M}_{\fs}(0,0)$ given by
\bd
\mathscr{M}_{\fs}(0,0) = \sqrt{\fa \fs^{2}} \lbk \vartheta_{3}(0 | 4 \nc \fa \fs^{-2}) \vartheta_{3}(0 | \nc \fa \fs^{2}) +
\vartheta_{2}(0 | 4 \nc \fa \fs^{-2}) \vartheta_{4}(0 | \nc \fa \fs^{2}) \rbk . 
\ed
In principle, the squeezing parameter $\fs$ is positive definite and assumes any values in the interval $\fs \in (0,\infty)$. Indeed,
the introduction of this factor in the second argument of the $\vartheta_{3}$-function allows us not only to modify the width of the
wavepacket (\ref{e7}) for values greater or smaller than $\sqrt{2 \fa}$, but also to generalize the theory of discrete representations
for finite-dimensional Hilbert spaces in order to include the discrete squeezed states within the scope of a general theory from which
the coherent states represent a particular case. Moreover, note that $\lgg u_{\gam} | 0; \fs \rg$ is a discrete representative of that
equivalence relation established in Ref. \cite{r22} between the wavepacket phase space representation and the phase space generated
by the continuous squeezed states, and represents the cornerstone of the present approach. Next, we will study two special classes of
discrete squeezed states in an $N^{2}$-dimensional phase space following the mathematical recipe described in the previous sections.

\subsection{Discrete representatives of the displaced squeezed-vacuum states}

Let us initially decompose the projector $| 0;\fs \rg \lgg 0;\fs |$ through the expansion formula (\ref{e2}) for $s=-1$ fixed. After
some algebraic manipulations, it is easy to show that
\be
\lb{e9}
| 0;\fs \rg \lgg 0;\fs | = \frac{1}{\sqrt{N}} \sum_{\eta,\xi = - \ell}^{\ell} \mathscr{K}_{\fs}(\eta,\xi) {\bf S}(\eta, \xi)
\ee
has a similar mathematical structure to that obtained for the projector of discrete coherent states -- here represented by the operator
${\bf T}^{(-1)}(0,0)$ -- where, in particular, $\mathscr{K}_{\fs}(\eta,\xi) \coloneq \mathscr{M}_{\fs}(\eta,\xi) / \mathscr{M}_{\fs}
(0,0)$ can be promptly determined with the help of the auxiliary function
\brr
\mathscr{M}_{\fs}(\eta,\xi) &=& \frac{\sqrt{\fa \fs^{2}}}{2} \Bigl[ \vartheta_{3} (\fa \eta | \nc \fa \fs^{2}) \vartheta_{3} (\fa \xi |
\nc \fa \fs^{-2}) + \vartheta_{3}(\fa \eta | \nc \fa \fs^{2}) \vartheta_{4} (\pi \fa \xi | \nc \fa \fs^{-2}) \mathrm{e}^{\nc \pi \eta}
\nn \\
& & + \, \vartheta_{4}(\fa \eta | \nc \fa \fs^{2}) \vartheta_{3} (\pi \fa \xi | \nc \fa \fs^{-2}) \mathrm{e}^{\nc \pi \xi} +
\vartheta_{4}(\fa \eta | \nc \fa \fs^{2}) \vartheta_{4}(\fa \xi | \nc \fa \fs^{-2}) \mathrm{e}^{\nc \pi (\eta + \xi + N)} \Bigr] . \nn
\err 
The next step consists in applying on both sides of equation (\ref{e9}) the generator $\sqrt{N} {\bf S}(\nu,-\mu)$ with the aim of
obtaining the discrete representatives of the displaced squeezed-vacuum states. Such procedure leads us to establish a closed-form
expression for $| \mu,\nu;\fs \rg \lgg \mu,\nu;\fs |$ by means of the identity \cite{r11}
\bd
\lbk \sqrt{N} {\bf S}(\nu,-\mu) \rbk {\bf S}(\eta,\xi) \lbk \sqrt{N} {\bf S}^{\dagger}(\nu,-\mu) \rbk = \exp \lbk - \frac{2 \pi \nc}{N}
( \eta \mu + \xi \nu) \rbk {\bf S}(\eta,\xi) ,
\ed
namely
\be
\lb{e10}
| \mu,\nu;\fs \rg \lgg \mu,\nu;\fs | = \frac{1}{\sqrt{N}} \!\! \sum_{\eta,\xi = - \ell}^{\ell} \exp \lbk - \frac{2 \pi\nc}{N} ( \eta
\mu + \xi \nu ) \rbk \mathscr{K}_{\fs}(\eta,\xi) {\bf S}(\eta,\xi) .
\ee
Hence, Eq. (\ref{e10}) can be considered as being the first analytic discrete expression of the displaced squeezed-vacuum states in a
finite phase-space. 

\begin{figure}[!t]
\centering
\begin{minipage}[b]{0.45\linewidth}
\includegraphics[width=4cm]{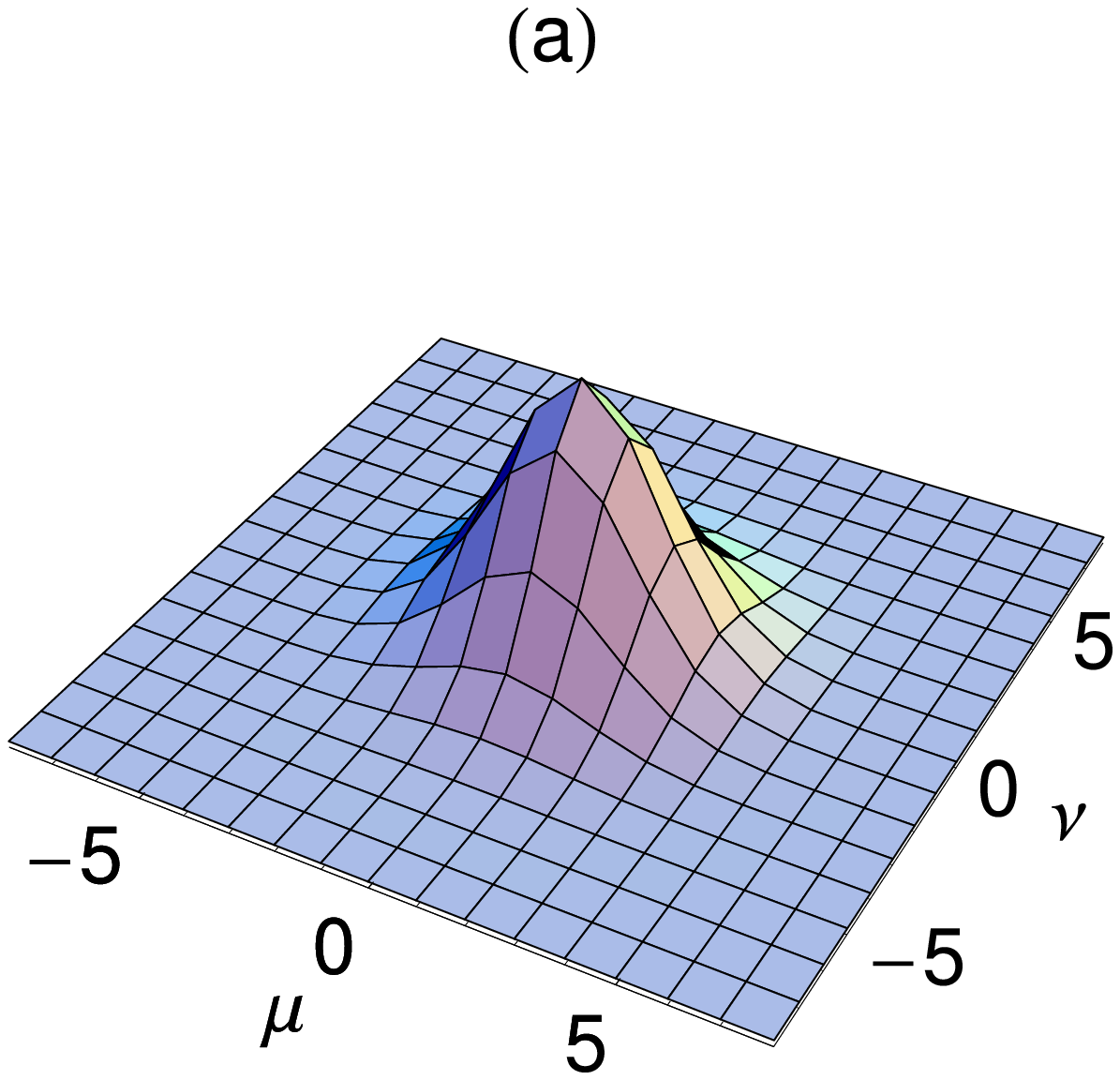}
\end{minipage} \hfill
\begin{minipage}[b]{0.45\linewidth}
\includegraphics[width=4cm]{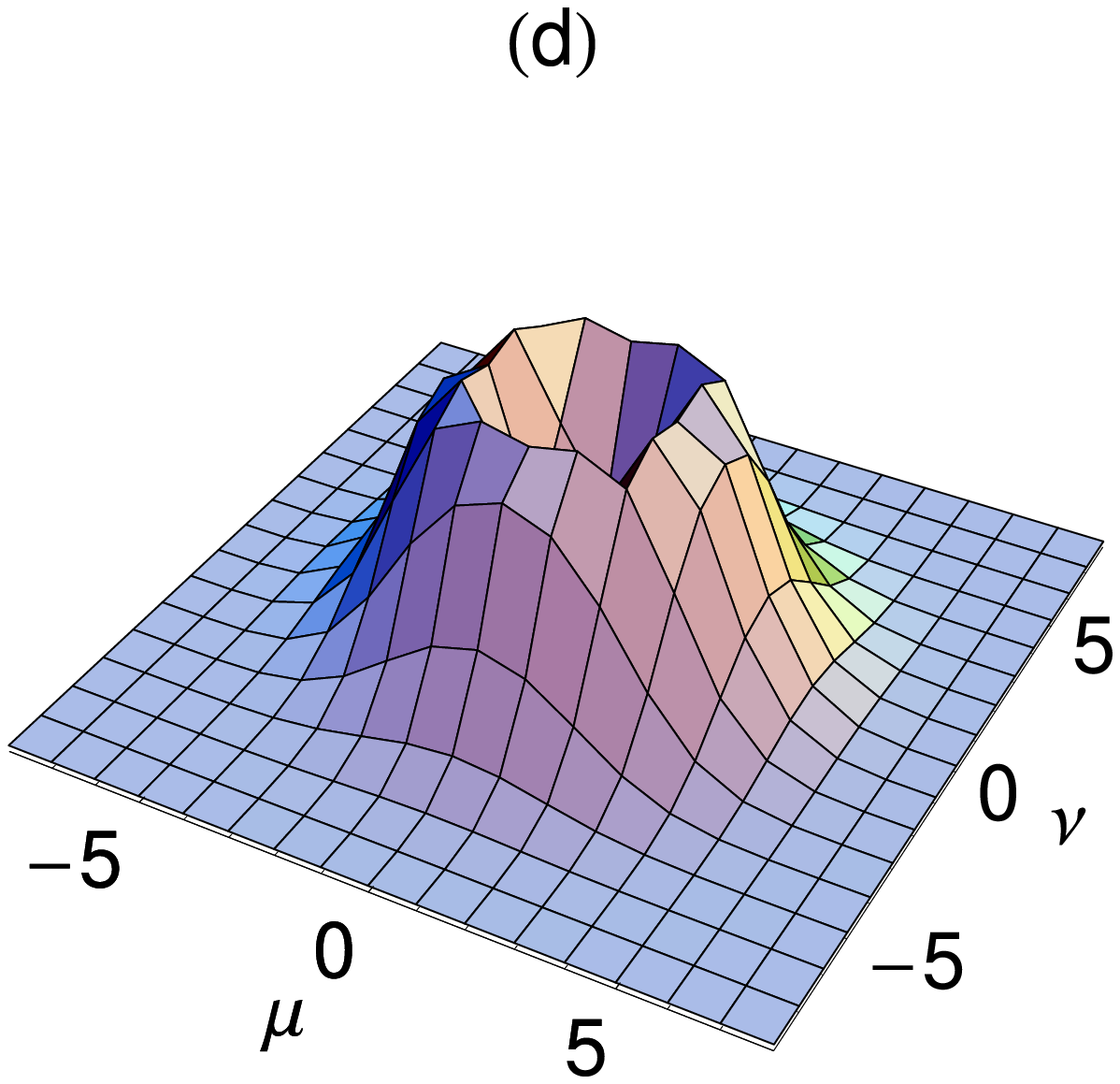}
\end{minipage} \hfill
\begin{minipage}[b]{0.45\linewidth}
\includegraphics[width=4cm]{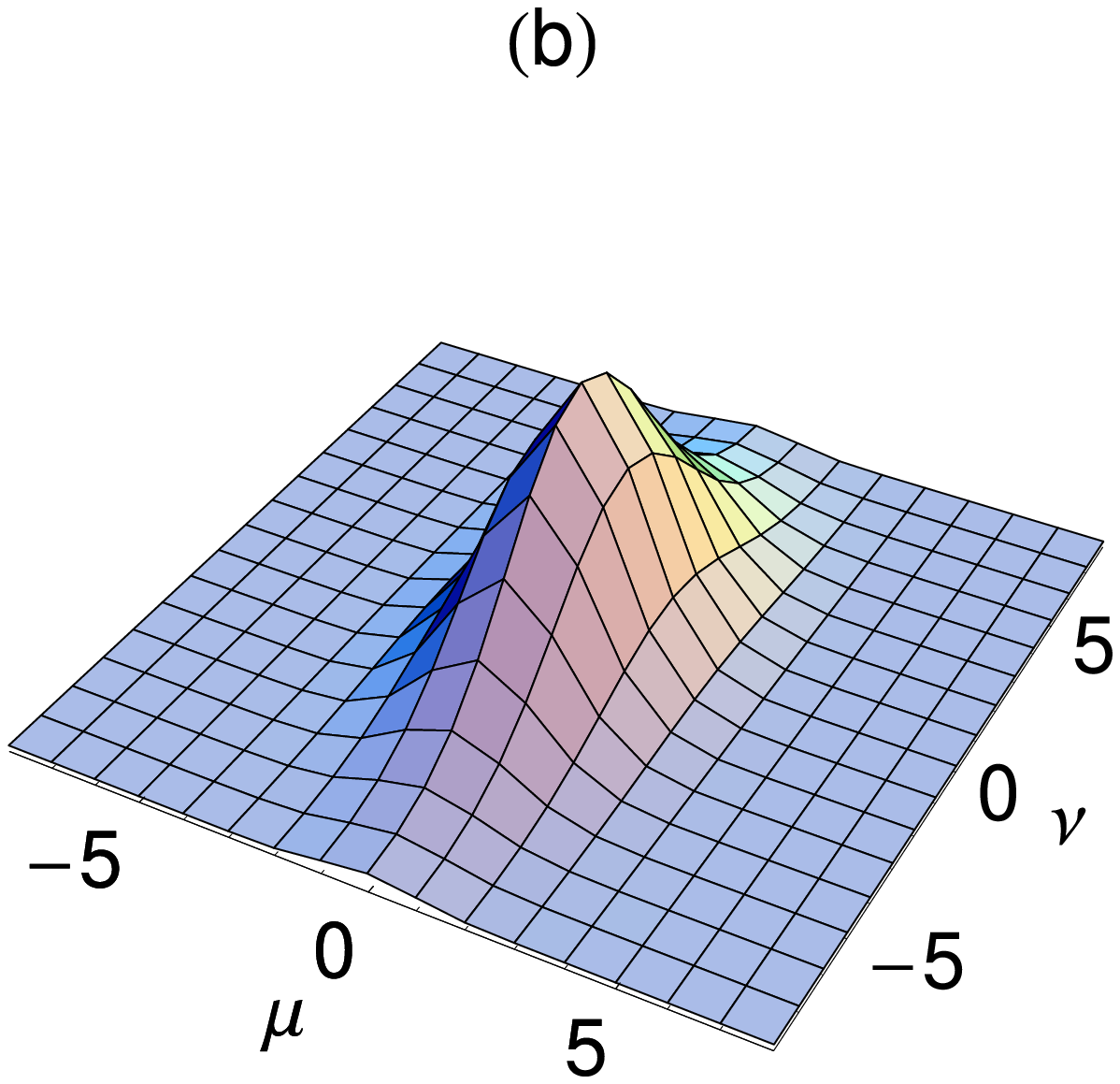}
\end{minipage} \hfill
\begin{minipage}[b]{0.45\linewidth}
\includegraphics[width=4cm]{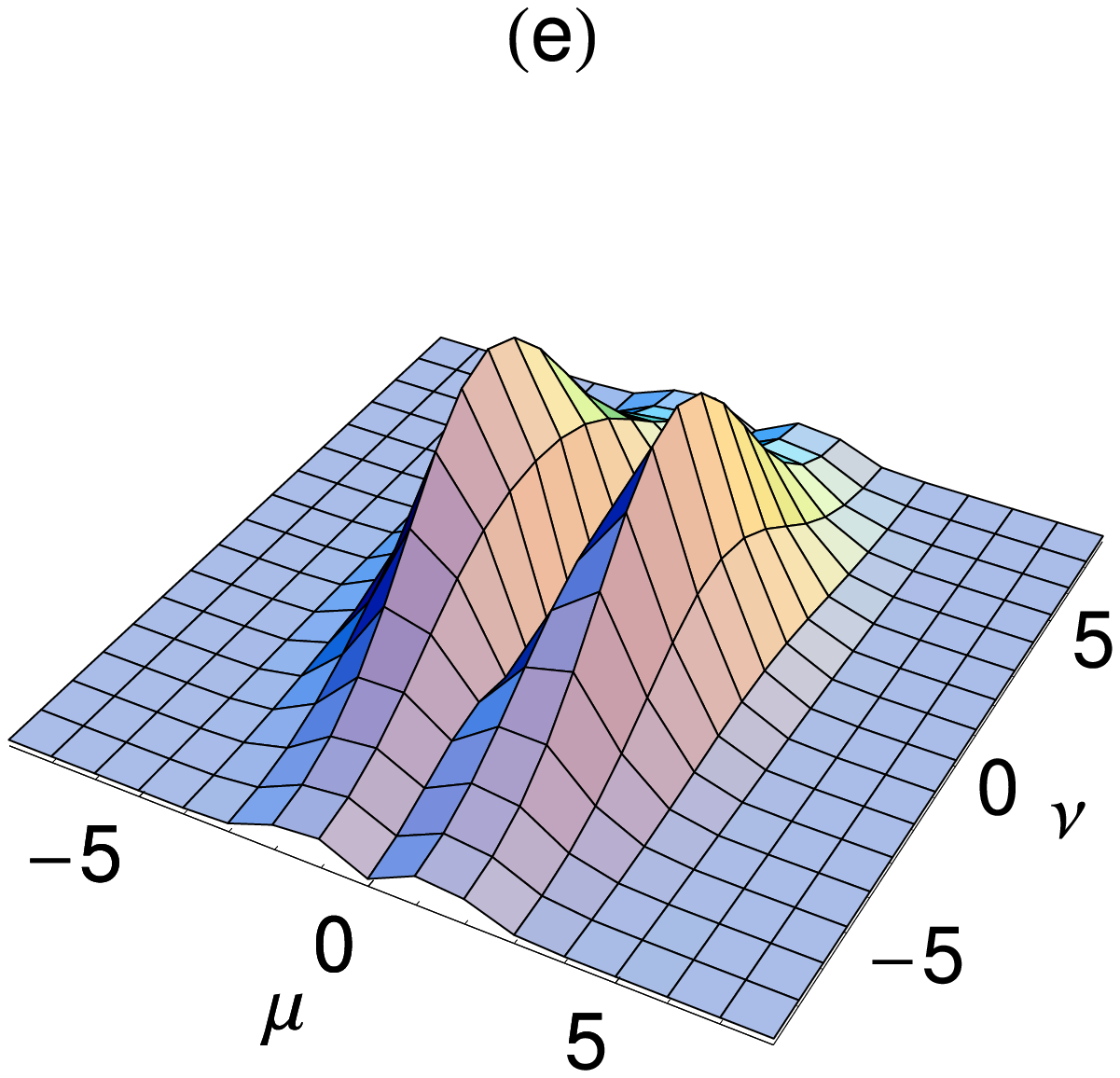}
\end{minipage} \hfill
\begin{minipage}[b]{0.45\linewidth}
\includegraphics[width=4cm]{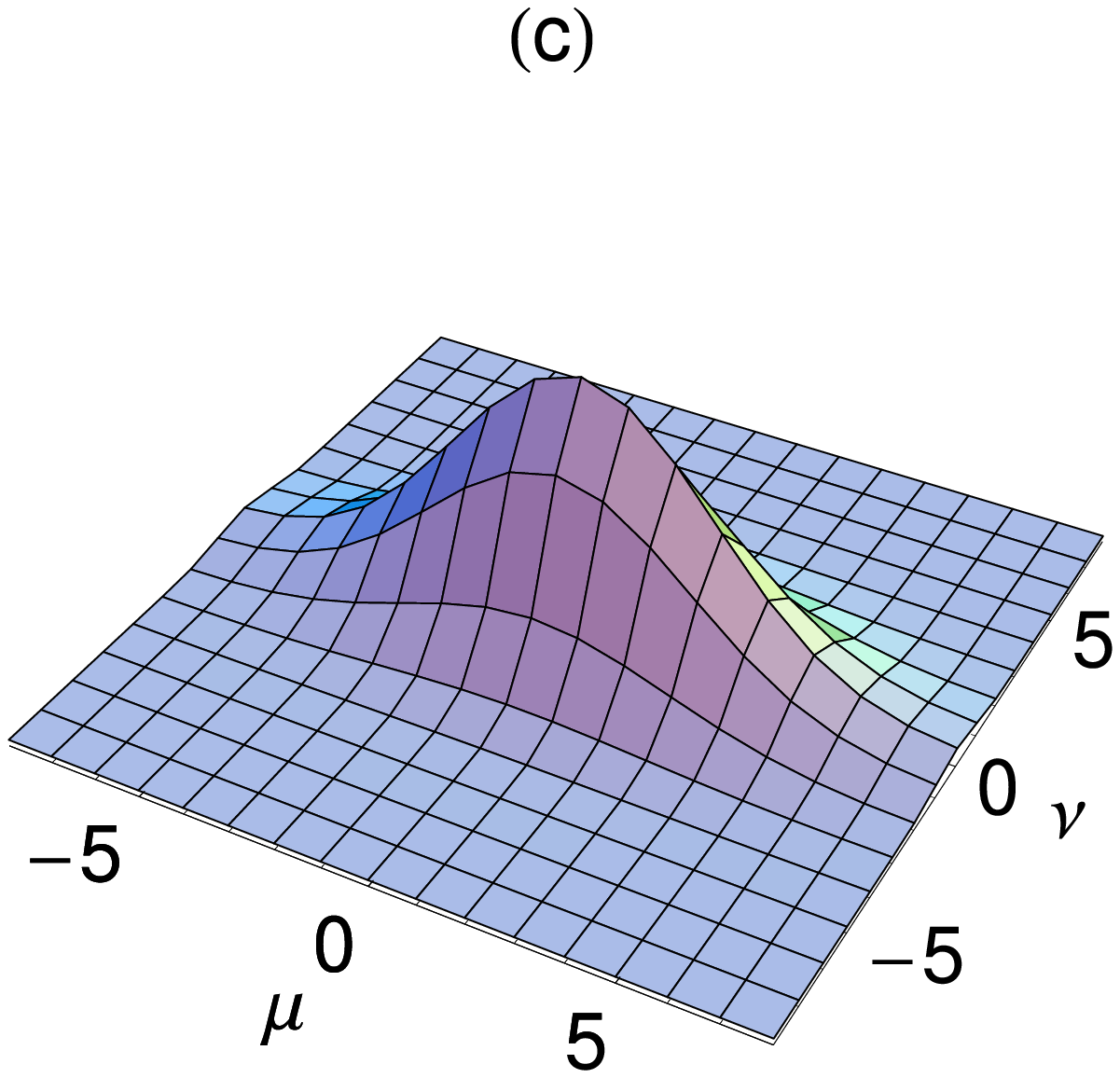}
\end{minipage} \hfill
\begin{minipage}[b]{0.45\linewidth}
\includegraphics[width=4cm]{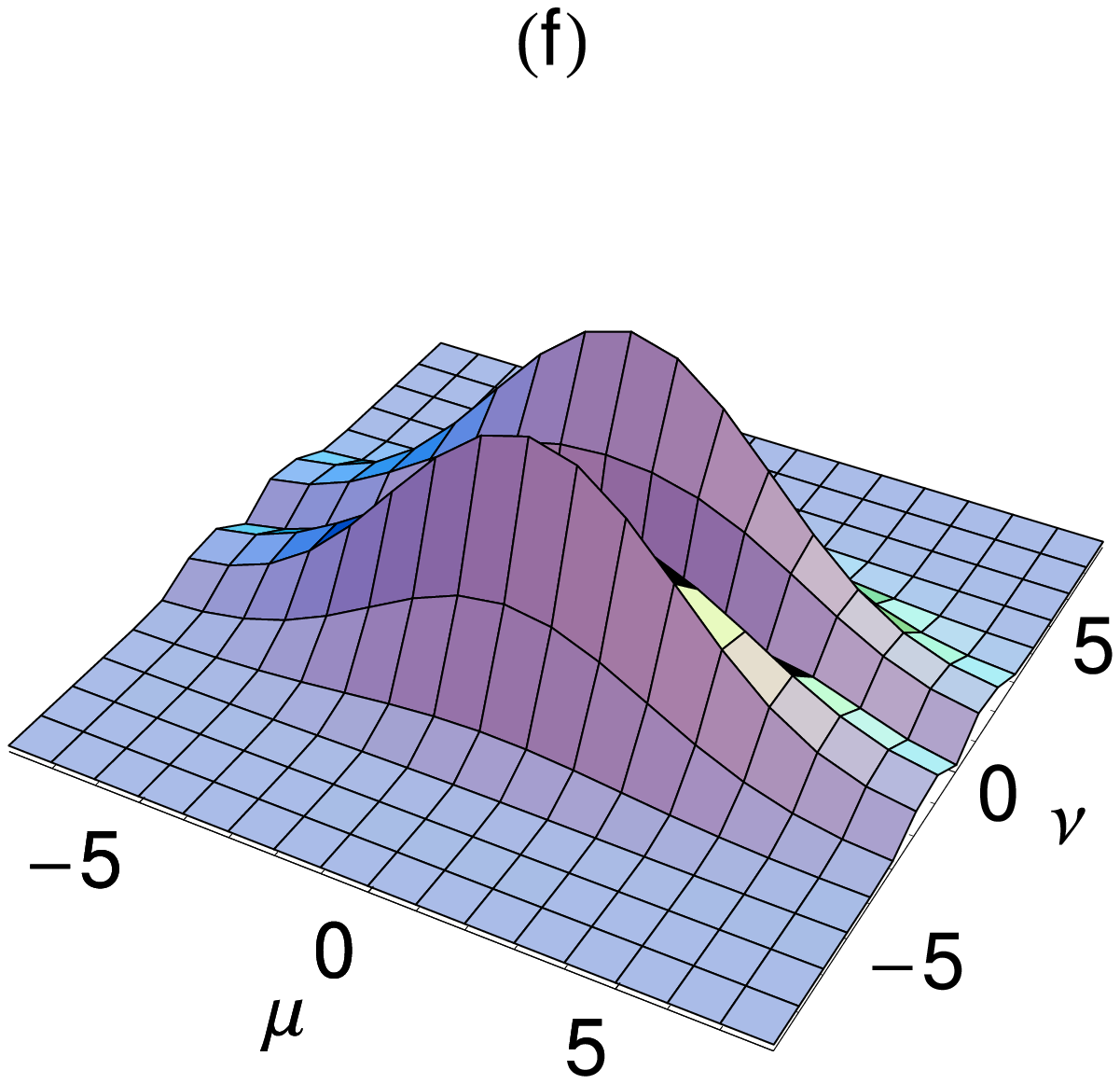}
\end{minipage}
\caption{Three-dimensional plots of (a,b,c) $P_{0}(\mu,\nu;\fs)$ and (d,e,f) $P_{1}(\mu,\nu;\fs)$ versus $\{ \mu,\nu \} \in [-8,8]$ for
$N=17$ fixed, and different values of squeezing parameter: (a,d) $\fs = 1$ (discrete coherent states), (b,e) $\fs^{2} = 5$ and (c,f)
$\fs^{-2} = 5$. Note that both the squeezing and stretching effects occur on the $17^{2}$-dimensional phase space labelled by the
dimensionless discrete variables $\mu$ and $\nu$. Indeed, these effects are directly linked to the modified second argument of the
Jacobi theta functions, which corroborate the approach here adopted for the discrete version of the displaced squeezed-vacuum states.} 
\end{figure}
An alternative form of Eq. (\ref{e10}) can also be found by expressing the product $N^{-1/2} \mathscr{K}(\eta,\xi) {\bf S}(\eta,\xi)$
in terms of a double Fourier transform of the diagonal projector $| \mu^{\prime},\nu^{\prime} \rg \lgg \mu^{\prime},\nu^{\prime} |$ for
the discrete coherent states, that is,
\be
\lb{e11}
| \mu,\nu;\fs \rg \lgg \mu,\nu;\fs | = \frac{1}{N^{2}} \sum_{\eta,\xi = - \ell}^{\ell} \exp \lbk - \frac{2 \pi \nc}{N} ( \eta \mu + \xi
\nu ) \rbk \mathds{K}(\eta,\xi;\fs) \sum_{\mu^{\prime},\nu^{\prime} = - \ell}^{\ell} \exp \lbk \frac{2 \pi \nc}{N} ( \mu^{\prime} \eta +
\nu^{\prime} \xi ) \rbk | \mu^{\prime},\nu^{\prime} \rg \lgg \mu^{\prime},\nu^{\prime} | 
\ee
where $\mathds{K}(\eta,\xi;\fs) \coloneq \mathscr{K}_{\fs}(\eta,\xi) / \mathscr{K}(\eta,\xi)$ denotes a new kernel-like function which
is responsible by the squeezing propa\-gation into a discrete finite-dimensional phase-space. An immediate application of this equation
refers to the evaluation of $| \lgg n | \mu,\nu;\fs \rg |^{2}$ through the results obtained until the present moment. For such task, it
is sufficient to calculate the overlap probability distribution $P_{n}(\mu,\nu;\fs)$ by means of the equation
\bd
P_{n}(\mu,\nu;\fs) = \frac{1}{N^{2}} \sum_{\eta,\xi = - \ell}^{\ell} \exp \lbk - \frac{2 \pi \nc}{N} ( \eta \mu + \xi \nu ) \rbk
\mathds{K}(\eta,\xi;\fs) \sum_{\mu^{\prime},\nu^{\prime} = - \ell}^{\ell} \exp \lbk \frac{2 \pi \nc}{N} ( \mu^{\prime} \eta +
\nu^{\prime} \xi ) \rbk P_{n}(\mu^{\prime},\nu^{\prime}) ,
\ed
which describes how the squeezing effect is propagated for a given initial distribution function $P_{n}(\mu^{\prime},\nu^{\prime})$.
Next, we substitute Eq. (\ref{e6}) into this expression and carry out the sums over $\mu^{\prime}$ and $\nu^{\prime}$ in order to
obtain the compact form
\be
\lb{e12}
P_{n}(\mu,\nu;\fs) = \frac{1}{N} \sum_{\eta,\xi = - \ell}^{\ell} \exp \lbk - \frac{2 \pi \nc}{N} ( \eta \mu + \xi \nu ) \rbk
\mathscr{K}_{\fs}(\eta,\xi) \mathscr{K}_{n}(\eta,\xi) .
\ee
This result states that $P_{n}(\mu,\nu;\fs)$ is promptly estimated through the double Fourier transform of the product
$\mathscr{K}_{\fs}(\eta,\xi) \mathscr{K}_{n}(\eta,\xi)$, and for $\fs = 1$ the overlap probability $P_{n}(\mu,\nu)$ can be immediately
recovered.

Figure 1 shows the three-dimensional plots of $P_{n}(\mu,\nu;\fs)$ versus $\{ \mu,\nu \} \in [-\ell,\ell]$ for (a,b,c) $n=0$ (vacuum
state) and (d,e,f) $n=1$ (first excited state) with $N=17$ fixed. In the numerical investigations, we have adopted some specific values
for $\fs$ in order to illustrate the squeezing and stretching effects within a finite-dimensional phase space characterized by the
discrete variables $\mu$ and $\nu$. For instance, pictures (a,d) correspond to the particular value $\fs=1$ (absence of squeezing
effect) and represent the three-dimensional plots of $P_{n}(\mu,\nu)$ -- see Eq. (\ref{e6}) for technical details -- when $n=0$ and
$1$, respectively. On the other hand, pictures (b,e) are related to the value $\fs^{2}=5$ and show some typical effects that confirm
the quantum assignment of the displaced squeezed-vacuum states. Furthermore, pictures (c,f) also exhibit similar structures for
$\fs^{-2}=5$, where now the squeezing (stretching) effect is promptly verified in the discrete variable $\nu$ $(\mu)$. In both cases,
the appearance of the squeezing and stretching effects can be understood in this context as a direct consequence of the construction
process adopted for the discrete representatives of the states under investigation. Indeed, the introduction of an extra parameter in
the second argument of the Jacobi theta functions has allowed us to properly construct out, from the finite-dimensional context, a wave
function that inherently embodies the discrete analogues of the desired properties of the continuous squeezed states. 

However, it is worth noticing that we have only discussed hitherto the action of a hypothetical squeezing generator on the discrete
vacuum state without deriving this operator and/or establishing its mathematical properties. This breach will be filled in the appendix
B through a discussion on the discrete representative of the squeezing generator for finite Hilbert spaces.

\subsection{Discrete representatives of the squeezed coherent states}

Initially, let us consider the decomposition of the diagonal projector related to the discrete coherent states $\{ | \bar{\mu},
\bar{\nu} \rg \}_{\bar{\mu},\bar{\nu} = - \ell,\ldots,\ell}$, that is
\be
\lb{e13}
| \bar{\mu},\bar{\nu} \rg \lgg \bar{\mu},\bar{\nu} | = \frac{1}{\sqrt{N}} \sum_{\eta,\xi = -\ell}^{\ell} \exp \lbk - \frac{2 \pi \nc}{N}
( \eta \bar{\mu} + \xi \bar{\nu} ) \rbk \mathscr{K}(\eta,\xi) {\bf S}(\eta,\xi) \; .
\ee
Since the discrete representatives of the squeezed coherent states are defined within this context through the action of the squeezing
operator $\sqq(\fs)$ on the discrete coherent states, the diagonal projector
\be
\lb{e14}
| \bar{\mu},\bar{\nu};\fs \rg \lgg \bar{\mu},\bar{\nu};\fs | = \sum_{\eta^{\prime},\xi^{\prime} = -\ell}^{\ell} \mathrm{Tr} \lbk 
{\bf S}^{\dagger}(\eta^{\prime},\xi^{\prime}) | \bar{\mu},\bar{\nu};\fs \rg \lgg \bar{\mu},\bar{\nu};\fs | \rbk {\bf S} (\eta^{\prime},
\xi^{\prime})
\ee
can be interpreted as being the decomposition of $\sqq(\fs) | \bar{\mu},\bar{\nu} \rg \lgg \bar{\mu},\bar{\nu} | \sqq^{\dag}(\fs)$ in a
complete orthonormal operator basis characterized by a set of $N^{2}$-operators $\{ {\bf S}(\eta^{\prime},\xi^{\prime}) \}$, where the
coefficients $\lgg \bar{\mu},\bar{\nu}; \fs | {\bf S}^{\dagger}(\eta^{\prime},\xi^{\prime}) | \bar{\mu},\bar{\nu};\fs \rg$ are
calculated through the formal expression
\be
\lb{e15}
N \sum_{m,n = 0}^{N-1} \Gamma_{0m}(-\bar{\nu},\bar{\mu};1) \Gamma_{mn}(-\eta^{\prime},-\xi^{\prime};\fs) 
\Gamma_{n0}(\bar{\nu},-\bar{\mu};1)
\ee
with the nondiagonal matrix elements in the finite squeezed-number basis $\Gamma_{mn}(\alf,\bet;\fs) \coloneq \lgg m;\fs | {\bf S}
(\alf,\bet) | n;\fs \rg$ defined as \cite{r18}
\bd
\Gamma_{mn}(\alf,\bet;\fs) = \exp \lpar - \frac{\nc \pi}{N} \alf \bet \rpar \sum_{\kappa = - \ell}^{\ell} \exp \lpar \frac{2 \pi \nc}{N}
\kappa \alf \rpar \mathfrak{F}_{\kappa-\bet,m}^{\ast}(\fs) \mathfrak{F}_{\kappa,n}(\fs) .
\ed
It is worth mentioning that these generalized nondiagonal matrix elements satisfy the relations $\Gamma_{mn}(0,0;\fs) =
\delta_{m,n}^{[\ns]}$ and $\Gamma_{00}(\alf,\bet;\fs)=\mathscr{K}_{\fs}(\alf,\bet)$, which are related to the orthogonality rule for the
finite squeezed-number states and the diagonal matrix element $\lgg 0;\fs | {\bf T}^{(s)}(\mu,\nu) | 0;\fs \rg$ for the squeezed-vacuum
state. Furthermore, for $m=n$ and $\fs=1$ it is possible to verify that $\Gamma_{nn}(\alf,\bet;1)$ coincides with $\mathscr{K}_{n}
(\alf,\bet)$. Next, we will derive an expression for the overlap probability distribution $\bar{P}_{n}(\bar{\mu},\bar{\nu};\fs)$ by 
means of an alternative form to that established in this paragraph for the diagonal projector (\ref{e14}).

The overlap probability distribution $\bar{P}_{n}(\bar{\mu},\bar{\nu};\fs) \coloneq | \lgg n | \bar{\mu},\bar{\nu};\fs \rg |^{2}$
associated with the discrete representatives of the squeezed coherent states can be evaluated directly from Eq. (\ref{e14}) as follows:
\be
\lb{e16}
\bar{P}_{n}(\bar{\mu},\bar{\nu};\fs) = \frac{1}{\sqrt{N}} \sum_{\eta^{\prime},\xi^{\prime} = -\ell}^{\ell} \lgg \bar{\mu},\bar{\nu};\fs
| {\bf S}^{\dagger}(\eta^{\prime},\xi^{\prime}) | \bar{\mu},\bar{\nu};\fs \rg \mathscr{K}_{n}(\eta^{\prime},\xi^{\prime}) \; .
\ee
On the other hand, if one considers the mathematical procedure used to derive Eq. (\ref{e11}) in the context here exposed, the
alternative expression
\brr
\lb{e17}
| \bar{\mu},\bar{\nu};\fs \rg \lgg \bar{\mu},\bar{\nu};\fs | &=& \frac{1}{N^{2}} \sum_{\eta^{\prime},\xi^{\prime} = - \ell}^{\ell}
\overbrace{\sqrt{N} \lbk \mathscr{K}(\eta^{\prime},\xi^{\prime}) \rbk^{-1} \lgg \bar{\mu},\bar{\nu};\fs | {\bf S}^{\dagger}
(\eta^{\prime},\xi^{\prime}) | \bar{\mu},\bar{\nu};\fs \rg}^{K(\bar{\mu}, \bar{\nu}, \eta^{\prime},\xi^{\prime};\fs)} \nn \\
& & \times \sum_{\mu^{\prime},\nu^{\prime} = -\ell}^{\ell} \exp \lbk \frac{2 \pi \nc}{N} ( \eta^{\prime} \mu^{\prime} + \xi^{\prime}
\nu^{\prime} ) \rbk | \mu^{\prime},\nu^{\prime} \rg \lgg \mu^{\prime},\nu^{\prime} |
\err
allows us to show that $\bar{P}_{n}(\bar{\mu},\bar{\nu};\fs)$ can be also estimated through a product of the kernel-like function
$K(\bar{\mu},\bar{\nu},\eta^{\prime},\xi^{\prime};\fs)$ -- which describes how the squeezing effect occurs in an $N^{2}$-dimensional
phase space labeled by the set of discrete variables $\{ \bar{\mu},\bar{\nu} \}$ -- and the double Fourier transform of the initial
distribution function $P_{n}(\mu^{\prime},\nu^{\prime})$,
\bd
\bar{P}_{n}(\bar{\mu},\bar{\nu};\fs) = \frac{1}{N^{2}} \!\! \sum_{\eta^{\prime},\xi^{\prime},\mu^{\prime},\nu^{\prime} = -\ell}^{\ell}
\!\!\! \exp \lbk \frac{2 \pi \nc}{N} ( \eta^{\prime} \mu^{\prime} + \xi^{\prime} \nu^{\prime} ) \! \rbk K(\bar{\mu}, \bar{\nu},
\eta^{\prime},\xi^{\prime};\fs) P_{n}(\mu^{\prime},\nu^{\prime}) . 
\ed
In the absence of squeezing effect, $K(\bar{\mu},\bar{\nu},\eta^{\prime},\xi^{\prime};1)$ coincides exactly with $\exp \lbk - (2 \pi \nc
/ N)(\eta^{\prime} \bar{\mu} + \xi^{\prime} \bar{\nu}) \rbk$, and consequently, the double sum over $\eta^{\prime}$ and $\xi^{\prime}$
leads us to reobtain, by means of the product of Kronecker delta functions $\delta_{\mu^{\prime},\bar{\mu}}^{[\ns]}
\delta_{\nu^{\prime},\bar{\nu}}^{[\ns]}$, the discrete probability distribution function (\ref{e6}). 

\section{Applications}

In Ref. \cite{r31} it has been mentioned the concrete possibility of using the degenerate parametric amplification of a signal to
reproduce, from the experimental point of view, the unitary transformation $\sop^{\dagger}(\xi) \ro \sop(\xi)$ generated by the
action of the squeezing operator upon physical systems described by continuous quantum variables \cite{r32}. However, if one deals
with specific physical systems characterized by quantum states of arbitrary finite dimension and described by discrete quantum
variables, the squeezing action will be modelled hypothetically through a nonlinear medium or other external apparatus connected to the
system of interest. This connection will then produce a global effect upon the system that consists in modifying its discrete
wavefunction according to the rules established by a particular squeezing operator. In this sense, let us suppose that $\sqq(\fs)$ can
be modelled by one of these hypotheses which leads one to produce, within certain experimental and theoretical limitations (e.g., the
degrading and ubiquitous decoherence due to the unavoidable coupling with the environment), the unitary transformation $\sqq^{\dagger}
(\fs) \ro \sqq(\fs)$ for physical systems belonging to finite-dimensional Hilbert spaces. Next, we will derive an analytical expression
for $F^{(s)}(\bar{\mu},\bar{\nu};\fs) \coloneq \mathrm{Tr} [ {\bf T}^{(s)}(\bar{\mu},\bar{\nu}) \sqq^{\dagger}(\fs) \ro \sqq(\fs) ]$
which allows us to obtain, as particular situations of this parametrized function, the discrete Husimi $(s=-1)$, Wigner $(s=0)$, and
Glauber-Sudarshan $(s=1)$ functions in an $N^{2}$-dimensional phase space.

For this purpose, let us initially consider the cyclic invariance property under trace operation in the definition of 
$F^{(s)}(\bar{\mu},\bar{\nu};\fs)$, namely 
\bd
F^{(s)}(\bar{\mu},\bar{\nu};\fs) = \mathrm{Tr} [ \sqq(\fs) {\bf T}^{(s)}(\bar{\mu},\bar{\nu}) \sqq^{\dagger}(\fs) \ro ] .
\ed
The next step consists in substituting Eq. (\ref{e1}) on the right-hand side of this equality. Thus, after some calculations, we
immediately recognize
\be
\lb{e18}
F^{(s)}(\bar{\mu},\bar{\nu};\fs) = \frac{1}{\sqrt{N}} \sum_{\eta,\xi = - \ell}^{\ell} \exp \lbk - \frac{2 \pi \nc}{N} (\eta \bar{\mu} +
\xi \bar{\nu}) \rbk \Xi^{(s)}(\eta,\xi;\fs)
\ee
as being the desired analytical expression for the parametrized function where, in particular,
\bd
\Xi^{(s)}(\eta,\xi;\fs) = \mathrm{Tr} [ \sqq(\fs) {\bf S}^{(s)}(\eta,\xi) \sqq^{\dagger}(\fs) \ro ]
\ed
represents the squeezed discrete $s$-ordered characteristic function. Note that for $\fs = 1$, the parametrized function
$F^{(s)}(\bar{\mu},\bar{\nu};1)$ coincides exactly with that obtained by Eq. (\ref{e4}). Now, we will concentrate our efforts in
describing a physical process which allows us to measure directly the discrete Wigner function $\mathcal{W}(\bar{\mu},\bar{\nu};\fs)$
for any quantum system constituted by a finite-dimensional state space.

\subsection{Measuring the discrete Wigner function via generalized scattering circuit}

Recently, we have employed a slightly modified version of the scattering circuit proposed by Paz and co-workers \cite{r23} to measure
the discrete Wigner function $\mathcal{W}(\mu,\nu)$ in the absence of squeezing effects \cite{r18}. In the present approach, we
generalize such version in order to introduce within the primary circuit an auxiliary set of programmable gate arrays which allows us
to reproduce the unitary transformation $\sqq^{\dagger}(\fs) \ro \sqq(\fs)$. This theoretical implementation will then open new
possibilities of measurements on the generalized circuit: for instance, both the discrete Wigner function $\mathcal{W}(\bar{\mu},
\bar{\nu};\fs)$ and the characteristic function $\Xi^{(0)}(\eta,\xi;\fs)$ will be inferred in the same experiment through a simple
operation associted with the controlled Fourier transform (FT). 

\begin{figure}[!t]
\centering
\begin{minipage}[b]{0.70\linewidth}
\includegraphics[width=0.8\textwidth]{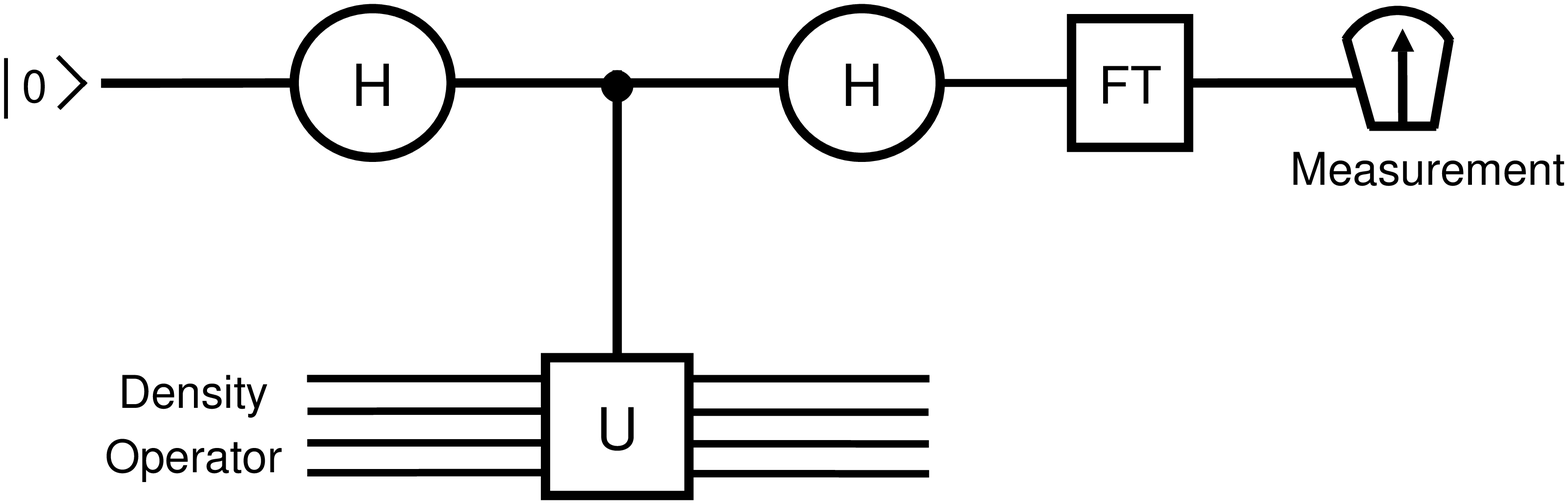}
\end{minipage}
\caption{Generalized version of the scattering circuit used to evaluate the real and imaginary parts of the expectation value
$\mathrm{Tr} [ {\bf U} \sqq^{\dagger}(\fs) \ro \sqq(\fs) ]$ for a unitary operator ${\bf U} = \sqrt{N} {\bf S}(\eta,\xi)$, where 
$| 0 \rg$ represents the ancillary qubit state which acts as a probe particle in a scattering experiment, and $\textrm{\bf H}$ denotes
a Hadamard transform. The measurements of the ancillary qubit polarizations along the $z$ and $y$ axes allow us to construct the
discrete Wigner function $\mathcal{W}(\bar{\mu},\bar{\nu};\fs) = \mathrm{Tr} [ {\bf T}^{(0)}(\bar{\mu},\bar{\nu}) \sqq^{\dagger}(\fs)
\ro \sqq(\fs) ]$ (discrete characteristic function $\Xi^{(0)}(\eta,\xi;\fs) = \mathrm{Tr} [ {\bf S}^{(0)}(\eta,\xi) \sqq^{\dagger}(\fs)
\ro \sqq(\fs) ]$) in the presence (absence) of the controlled-FT operation.}
\end{figure}
Following, let us describe some important implementations introduced within the scattering circuit which is illustrated in Fig. 2.
Basically, we modify this circuit by inserting a controlled-${\bf U}$ operation between the Hadamard gates, with ${\bf U} = \sqrt{N} \,
{\bf S}(\eta,\xi)$ acting on a squeezed quantum system whose density operator is exactly described by $\sqq^{\dagger}(\fs) \ro 
\sqq(\fs)$, and also a controlled-FT operation after the second Hadamard gate. This procedure allows us to obtain a set of measurements
on the polarizations along the $z$ and $y$ axes of the ancillary qubit $| 0 \rg$ that leads to the expectation values $\lgg \sig_{z}
\rg = \sqrt{N} \, \mathrm{Re} [ \mathcal{W}(\bar{\mu},\bar{\nu};\fs) ]$ and $\lgg \sig_{y} \rg = \sqrt{N} \, \mathrm{Im} [ \mathcal{W}
(\bar{\mu},\bar{\nu};\fs) ]$, respectively. Furthermore, in the absence of the controlled-FT operation, these measurements permit us to
reach the squeezed characteristic function $\Xi^{(s)}(\eta,\xi;\fs)$ for $s=0$ -- \ie, $\lgg \sig_{z} \rg = \sqrt{N} \, \mathrm{Re}
[ \Xi^{(0)}(\eta,\xi;\fs) ]$ and $\lgg \sig_{y} \rg = \sqrt{N} \, \mathrm{Im} [ \Xi^{(0)}(\eta,\xi;\fs) ]$. 

\subsection{A theoretical study on the quantum interference effects}

To conclude this section, we will present a theoretical study on the quantum interference effects between the discrete variables of an
$N^{2}$-dimensional phase space. Indeed, the concept of information entropy for the continuous squeezed states \cite{r33} will be
extended in order to include a consistent mathematical formulation based on the discrete squeezed states. Hence, let us introduce the
entropy functional
\be
\lb{e19}
\mathrm{E}[\mathcal{H};\fs] \coloneq - \frac{1}{N} \sum_{\mu,\nu = - \ell}^{\ell} \mathcal{H}(\mu,\nu;\fs) \ln \lbk \mathcal{H}
(\mu,\nu;\fs) \rbk \; ,
\ee
where $\mathcal{H}(\mu,\nu;\fs) \coloneq \lgg \mu,\nu;\fs | \ro | \mu,\nu;\fs \rg$ represents the discrete Husimi function related to
the displaced squeezed-vacuum states. The virtues of this definition are directly associated with its continuous counterpart, that is,
(i) once $\mathcal{H}(\mu,\nu;\fs)$ is a positive definite probability distribution function and limited to the interval $[0,1]$, the
joint entropy $\mathrm{E}[\mathcal{H};\fs]$ characterizes a well-defined function whose behaviour does not present any mathematical
inconsistencies. Besides, (ii) this joint entropy essentially measures the functional correlation between the discrete eigenvalues of
coordinate and momentum-like operators for a specific class of quantum systems with finite Hilbert spaces. Finally, it is worth
mentioning that $\mathrm{E}[\mathcal{H};\fs]$ can also be defined for the discrete representatives of the squeezed coherent states or
any discrete representation related to a finite phase-space.

Next, let us consider the functionals related to the marginal entropies
\be
\lb{e20}
\mathrm{E}[\mathcal{Q};\fs] \coloneq - \frac{1}{\sqrt{N}} \sum_{\mu = - \ell}^{\ell} \mathcal{Q}(\mu;\fs) \ln \lbk \mathcal{Q}(\mu;\fs)
\rbk 
\ee
and
\be
\lb{e21}
\mathrm{E}[\mathcal{R};\fs] \coloneq - \frac{1}{\sqrt{N}} \sum_{\nu = - \ell}^{\ell} \mathcal{R}(\nu;\fs) \ln \lbk \mathcal{R}(\nu;\fs)
\rbk ,
\ee
which depend on the discrete marginal distributions \cite{r18}
\brr
\mathcal{Q}(\mu;\fs) &=& \sum_{\eta = - \ell}^{\ell} \exp \lpar - \frac{2 \pi \nc}{N} \eta \mu \rpar
\frac{\mathscr{K}_{\fs}(\eta,0)}{\mathscr{K}(\eta,0)} \, \Xi^{(-1)}(\eta,0) \; , \nn \\
\mathcal{R}(\nu;\fs) &=& \sum_{\xi = - \ell}^{\ell} \exp \lpar - \frac{2 \pi \nc}{N} \xi \nu \rpar
\frac{\mathscr{K}_{\fs}(0,\xi)}{\mathscr{K}(0,\xi)} \, \Xi^{(-1)}(0,\xi) \; . \nn
\err
Consequently, the conditional entropies are given by
\be
\lb{e22}
\mathrm{E}[\mathcal{H} / \mathcal{Q};\fs] \coloneq - \frac{1}{N} \sum_{\mu,\nu = - \ell}^{\ell} \mathcal{H}(\mu,\nu;\fs) \ln \lbk
\frac{\mathcal{H}(\mu,\nu;\fs)}{\mathcal{Q}(\mu;\fs)} \rbk = \mathrm{E}[\mathcal{H};\fs] - \mathrm{E}[\mathcal{Q};\fs]
\ee
and
\be
\lb{e23}
\mathrm{E}[\mathcal{H} / \mathcal{R};\fs] \coloneq - \frac{1}{N} \sum_{\mu,\nu = - \ell}^{\ell} \mathcal{H}(\mu,\nu;\fs) \ln \lbk
\frac{\mathcal{H}(\mu,\nu;\fs)}{\mathcal{R}(\nu;\fs)} \rbk = \mathrm{E}[\mathcal{H};\fs] - \mathrm{E}[\mathcal{R};\fs] .
\ee
Note that (\ref{e22}) and (\ref{e23}) allow us not only to establish the balance equation
\be
\lb{e24}
\mathrm{E}[\mathcal{H} / \mathcal{Q};\fs] + \mathrm{E}[\mathcal{Q};\fs] = \mathrm{E}[\mathcal{H} / \mathcal{R};\fs] + \mathrm{E}
[\mathcal{R};\fs] ,
\ee
but also to derive some mathematical relations among these functionals from the Araki-Lieb inequality \cite{r34} 
\be
\lb{e25}
\left| \mathrm{E}[\mathcal{Q};\fs] - \mathrm{E}[\mathcal{R};\fs] \right| \leq \mathrm{E}[\mathcal{H};\fs] \leq \mathrm{E}
[\mathcal{Q};\fs] + \mathrm{E}[\mathcal{R};\fs] .
\ee
For instance, it is easy to show that the inequalities $\mathrm{E}[\mathcal{H} / \mathcal{Q};\fs] \leq \mathrm{E}[\mathcal{R};\fs]$ and
$\mathrm{E}[\mathcal{H} / \mathcal{R};\fs] \leq \mathrm{E}[\mathcal{Q};\fs]$ can be promptly obtained from the definitions established
for the conditional entropies and the right-hand side of Eq. (\ref{e25}) -- which defines the subadditivity property. Moreover, the
equal signs hold in both situations only when the discrete variables $\{ \mu,\nu \} \in [-\ell,\ell]$ are functionally uncorrelated,
namely, $\mathcal{H}(\mu,\nu;\fs)$ factorizes in the product $\mathcal{Q}(\mu;\fs) \mathcal{R}(\nu;\fs)$.

\begin{figure}[!th]
\centering
\begin{minipage}[b]{0.3\linewidth}
\includegraphics[width=4cm]{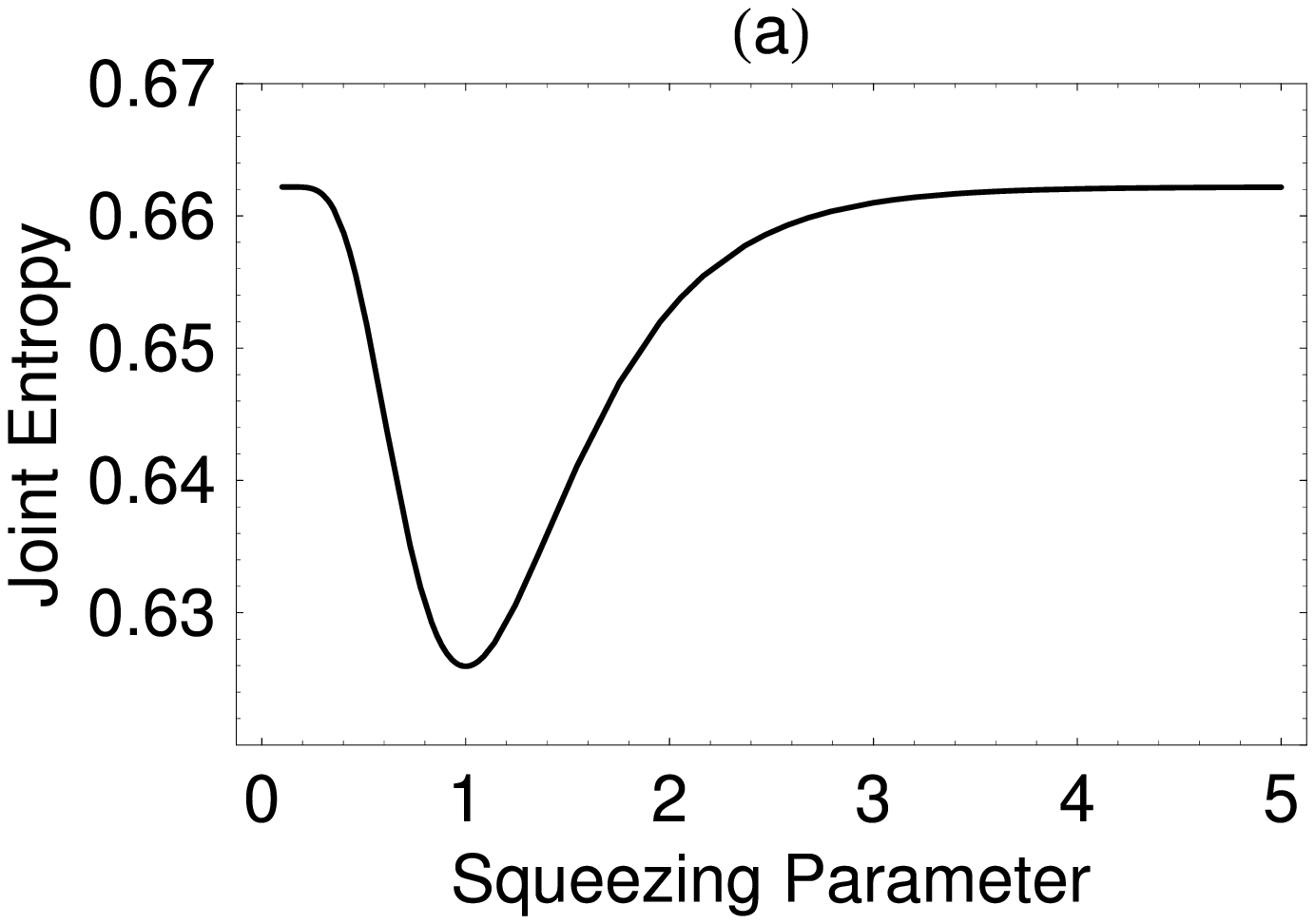}
\end{minipage} \hfill
\begin{minipage}[b]{0.3\linewidth}
\includegraphics[width=4cm]{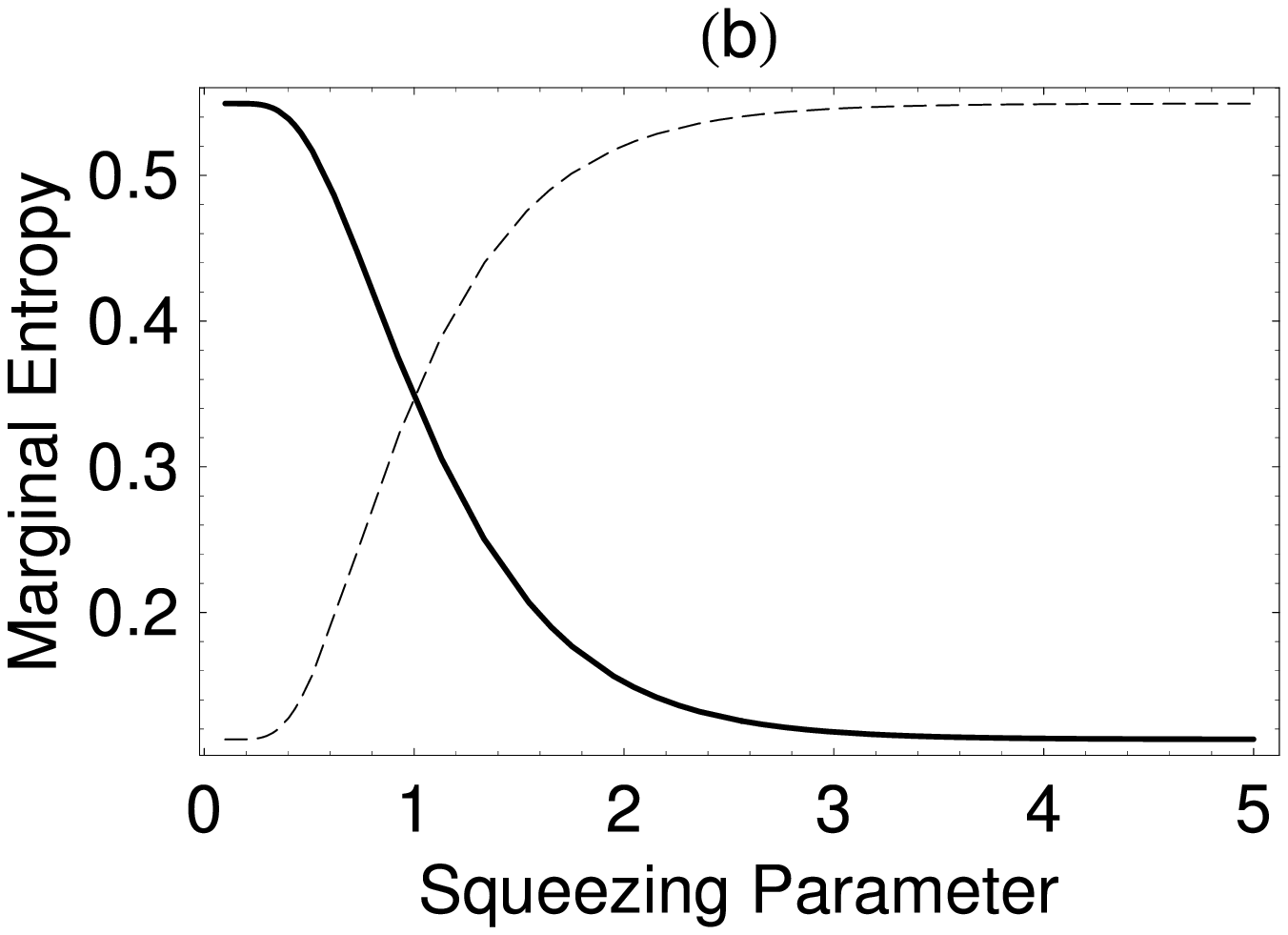}
\end{minipage} \hfill
\begin{minipage}[b]{0.3\linewidth}
\includegraphics[width=4cm]{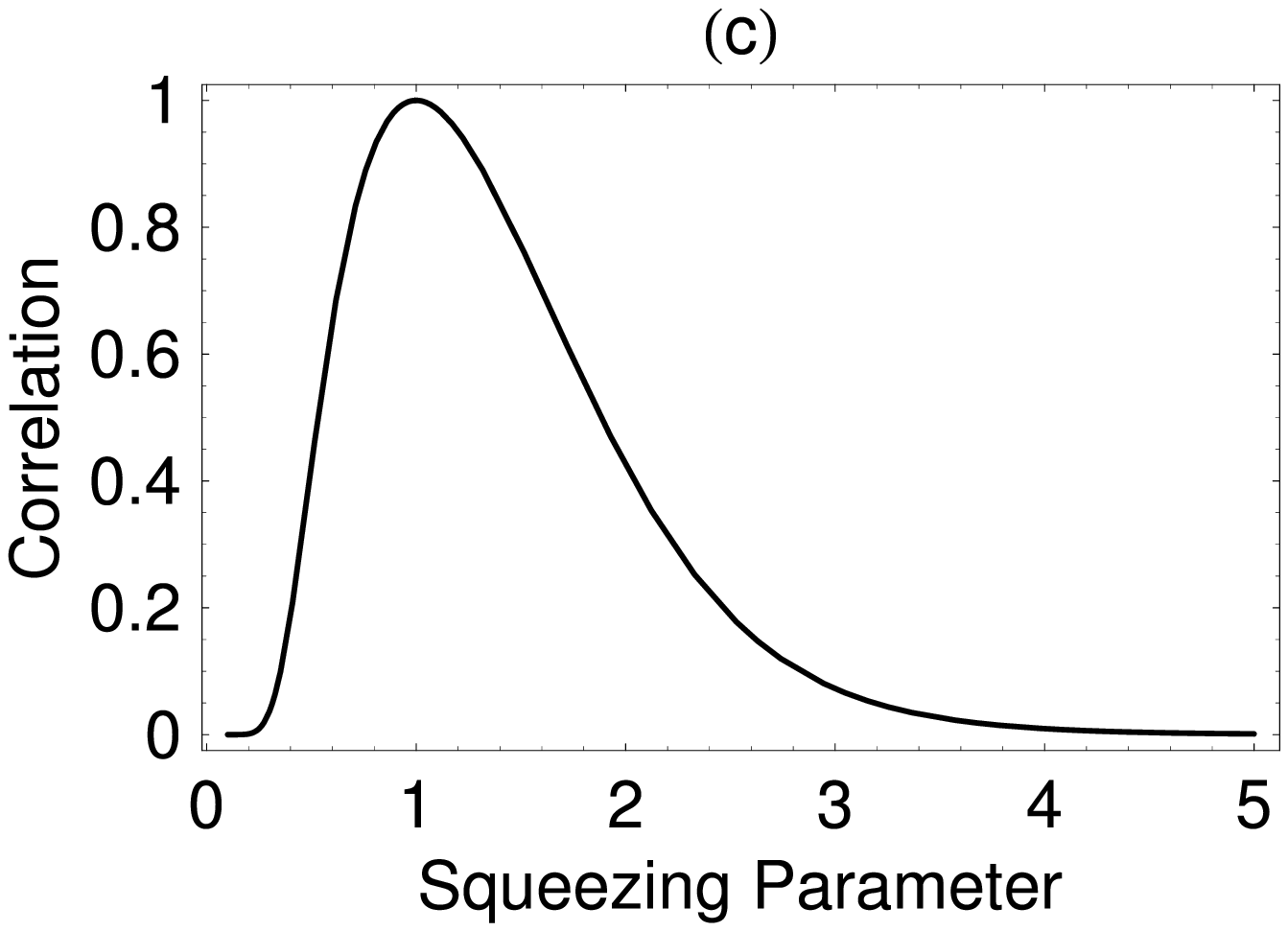}
\end{minipage}
\caption{Plots of some entropy functionals associated with the lowest level of the finite number state (vacuum state) as function
of the dimensionless squeezing parameter $\fs$ for $N=3$ fixed. Picture (a) shows the behaviour of the joint entropy
$\mathrm{E}[P_{0};\fs]$ related to the discrete Husimi distribution function $P_{0}(\mu,\nu;\fs)$, where its minimum value for 
$\fs = 1$ is approximately given by $0.625948$. The solid and dashed lines represent in picture (b) a graphical illustration of the
marginal entropies $\mathrm{E}[\mathcal{Q}_{0};\fs]$ and $\mathrm{E}[\mathcal{R}_{0};\fs]$, respectively. Note that both marginal
entropies present a distinct behaviour when $\fs \rightarrow 0$ or $\fs \gg 1$, and this fact implies in the absence of knowledge about
one of the discrete variables $\mu$ or $\nu$. The correlation functional is then plotted in picture (c) and its maximum value, attained
in $\fs = 1$, corroborates the numerical results obtained in the previous figures.}
\end{figure}
To avoid any ambiguity in the significance of the subadditivity property and to clarify its precise meaning, it is useful at this moment
to define the correlation functional
\be
\lb{e26}
\mathrm{C}[\mathcal{H};\fs] \coloneq \frac{\mathrm{E}[\mathcal{Q};\fs] + \mathrm{E}[\mathcal{R};\fs] - \mathrm{E}[\mathcal{H};\fs]}
{\mathrm{E}[\mathcal{Q};\fs] + \mathrm{E}[\mathcal{R};\fs] - \mathrm{Min}(\mathrm{E} [\mathcal{H};\fs])} ,
\ee
where $\mathrm{Min}(\mathrm{E}[\mathcal{H};\fs])$ denotes the global minimum of $\mathrm{E}[\mathcal{H};\fs]$ for a given dimension of
the Hilbert space and any squeezing parameter \cite{r35}. Essentially, $\mathrm{C}[\mathcal{H};\fs]$ measures the strengh of the
functional correlation between the discrete variables $\mu$ and $\nu$, having as reference a factorizable Husimi function
$\mathcal{H}(\mu,\nu;\fs)$. In addition, Eq. (\ref{e26}) assumes any values restricted to the interval $[0,1]$ such that its inferior
limit corresponds to the situation where the joint distribution function factorizes, while the superior limit is associated with the
opposite situation -- \ie, when the discrete variables are completely correlated. Hence, $\mathrm{C} [\mathcal{H};\fs]$ can be
considered as a first reasonable measure of functional correlation between the discrete variables of an $N^{2}$-dimensional phase
space, being its extension to multipartite systems very useful within the context of quantum information theory \cite{r24} and/or
other possible scenarios in physics \cite{r36}.

As a first practical application for these entropy functionals let us consider the discrete Husimi distribution function
$P_{0}(\mu,\nu;\fs)$ associated with the lowest level of the finite number state (namely, the vacuum state). Figure 3 shows the
plots of (a) joint entropy, (b) marginal entropies, and (c) correlation as function of the squeezing parameter for $N=3$ fixed. In
particular, we verify through numerical investigations that the minimum value attained by $\mathrm{E}[P_{0};\fs]$ in picture (a) for
$\fs = 1$ depends essentially on the Hilbert-space dimensionality. To illustrate this kind of dependence, we have evaluated the joint
entropy $\mathrm{E}[P_{0};1]$ for $N=3,5,7,9$ and obtained, as final results, the approximated values 0.625948, 0.953965, 0.992272, and
0.998598, respectively. Furthermore, the asymptotic value $\mathrm{Min}(\mathrm{E}[P_{0};1]) \rightarrow 1$ may be approximatelly
reached for $N \geq 11$, this value being coincident with that found for the continuous analogue \cite{r33}. On the other hand, it is
worth noticing that $\mathrm{E}[\mathcal{Q}_{0};\fs]$ (solid line) and $\mathrm{E}[\mathcal{R}_{0};\fs]$ (dashed line) in picture (b)
have a different behaviour when $\fs \rightarrow 0$ or $\fs \gg 1$, and this fact can be explained by means of the absence of knowledge
(uncertainty) about the discrete coordinate and momentum-like variables for each specific situation of $\fs$. Since the minimum value
of $\mathrm{E}[\mathcal{P}_{0};\fs]$ implies in maximum information gain on both the discrete variables of a finite-dimensional phase
space, it is expected that $\mathrm{C}[P_{0};\fs]$ attains a maximum value in this case (indeed, this is what exactly happens in picture
(c) for $\fs = 1$). However, if one considers the continuum limit $N \rightarrow \infty$, it is easy to show that $P_{0}(q,p;\lam) =
Q_{0}(q;\lam) R_{0}(p;\lam)$, \ie, the Husimi distribution function is completely uncorrelated, and consequently, the correlation
functional becomes zero for any value of squeezing parameter. 

\section{Conclusions}

In this paper we have established a set of important formal results that allows us to obtain a discrete analog for two well-known
definitions of squeezed states in a continuous phase-space \cite{r37,r38}. In fact, we have employed the mathematical fundamentals
developed in Ref. \cite{r18} with the aim of constructing out a consistent formalism for the discrete representatives of the displaced
squeezed-vacuum states and squeezed coherent states through their diagonal projectors in an $N^{2}$-dimensional phase space. Next, we
have applied our formalism to the context of quantum tomography and quantum information theory in order to obtain a finite phase-space
description of an important topic in physics which was nominated in the recent past as quantum interference effect \cite{r39}. It is
worth noticing that such description has allowed us to attain new results, within which some of them deserve to be mentioned: (i) we
have introduced a generalized version (where the squeezing effects are taking into account) of the scattering circuit proposed by Paz
and co-workers \cite{r23} to measure the discrete Wigner function or its associated characteristic function for any physical system
described by an $N$-dimensional space of states; (ii) we have also established a first reasonable measure of functional correlation
between the discrete variables of a finite phase-space whose extension to quantum information theory, as a measure of entanglement in
multipartite systems, is completely feasible; and finally, (iii) we have performed numerical investigations involving the discrete
squeezed-vacuum state and showed that, in particular, the maximum value of this measure presents a strong dependence on the
Hilbert-space dimension. As a concluding remark, it is worth mentioning that the formalism developed here opens new possibilities of
investigation mainly in modern research on quantum optics as well as the foundations of quantum mechanics. These considerations are
under current research and will be published elsewhere.

\section*{Acknowledgements}

MAM and DG (partially) are supported by Conselho Nacional de Desenvolvimento Cient\'{\i}fico e Tecnol\'ogico (CNPq), Brazil. The authors
are grateful to R. J. Napolitano for reading the manuscript and for providing valuable suggestions.

\appendix
\section{Calculational details of the diagonal matrix element $\lgg n| {\bf S}(\eta,\xi) |n \rg$}

In this appendix, we will explore the technique of breaking infinite sums in mod($N$) equivalence classes to derive a closed-form
analytical expression for the diagonal matrix element $\mathscr{K}_{n}(\eta,\xi) = \sqrt{N} \, \lgg n | {\bf S}(\eta,\xi) |n \rg$, with
\be
\lb{a1}
\mathscr{K}_{n}(\eta,\xi) = \exp \lpar - \frac{\nc \pi}{N} \eta \xi \rpar \sum_{\sigma = - \ell}^{\ell} \exp \lpar \frac{2 \pi \nc}{N}
\sigma \eta \rpar \mathfrak{F}_{\sigma,n} \, \mathfrak{F}^{\ast}_{\sigma - \xi, n}
\ee
written in terms of the coefficients \cite{r11,r26}
\brr
\lb{a2}
\mathfrak{F}_{\kappa,n} &=& \mathrm{N}_{n} \frac{(-\nc)^{n}}{\sqrt{N}} \sum_{\bet \in \mathbb{Z}} \exp \lpar - \frac{\pi}{N} \bet^{2} +
\frac{2 \pi \nc}{N} \bet \kappa \rpar H_{n} \lpar \sqrt{\frac{2 \pi}{N}} \, \bet \rpar \nn \\
&=& \frac{\mathrm{N}_{n}}{\sqrt{( \pi \fa )^{n} N}} \left. \frac{\upartial^{n}}{\upartial \fz^{n}} \exp (\pi \fa \fz^{2}) \,
\vartheta_{3} \lbk 2 \fa ( \kappa - \fz ) | 2 \nc \fa \rbk \right|_{\fz = 0} \; ,
\err
where $\mathrm{N}_{n}$ is the normalization constant and $H_{n}(z)$ is a Hermite polynomial. Basically, our goal will be split up into
two parts: the first one corresponds to the evaluation of the normalization constant $\mathrm{N}_{n}$, while the second one is
associated with the derivation of the generalized function $\mathscr{K}_{n}(\eta,\xi)$. 

\subsection{Normalization constant}

The constant $\mathrm{N}_{n}$ can be evaluated directly from the normalization relation for the finite number basis 
$\{ | n \rg \}_{n=0,\ldots,N-1}$, namely
\be
\lb{a3}
1 = \lgg n | n \rg = \sum_{\kappa = - \ell}^{\ell} \lgg n | u_{\kappa} \rg \lgg u_{\kappa} | n \rg = \sum_{\kappa = - \ell}^{\ell}
\mathfrak{F}^{\ast}_{\kappa,n} \, \mathfrak{F}_{\kappa,n} ,
\ee
where $\{ | u_{\kappa} \rg \}_{\kappa = - \ell,\ldots,\ell}$ represent the eigenstates of the Schwinger unitary operator ${\bf U}$.
Then, substituting the expression for $\mathfrak{F}_{\kappa,n}$ in Eq. (\ref{a3}) we obtain, as a first step in the calculations, the
equality
\bd
1 = \frac{\mathrm{N}_{n}^{2}}{N} \sum_{\kappa = - \ell}^{\ell} \, \sum_{\{ \alf,\bet \} \in \mathbb{Z}} \exp \lbk - \frac{\pi}{N}
(\alf^{2} + \bet^{2}) + \frac{2 \pi \nc}{N} \kappa (\bet - \alf) \rbk H_{n} \lpar \sqrt{\frac{2 \pi}{N}} \, \alf \rpar H_{n} \lpar
\sqrt{\frac{2 \pi}{N}} \, \bet \rpar . 
\ed
At this point, it is worth noticing that the sum over $\kappa$ can be readily carried out, which gives us 
$N \delta_{\alf,\bet}^{[\ns]}$ (the superscript $[N]$ on the Kronecker delta denotes that this function is different from zero when its
labels are congruent modulo $N$). Thus, $\alf$ will assume the values $\bet + \bet^{\prime} N$, with arbitrary $\bet^{\prime} \in
\mathbb{Z}$, yielding
\bd
1 = \mathrm{N}_{n}^{2} \sum_{\{ \bet,\bet^{\prime} \} \in \mathbb{Z}} \exp \lbr - \frac{\pi}{N} \lbk \bet^{2} + (\bet + \bet^{\prime}
N)^{2} \rbk \rbr H_{n} \lpar \sqrt{\frac{2 \pi}{N}} \, \bet \rpar H_{n} \lbk \sqrt{\frac{2 \pi}{N}} \, (\bet + \bet^{\prime} N) 
\rbk . 
\ed

The second step consists in considering the generating function for the product of two Hermite polynomials \cite{ap1}
\bd
H_{n}(x) H_{n}(y) = \left. \frac{\upartial^{n}}{\upartial \fz^{n}} ( 1-4 \fz^{2} )^{-1/2} \exp \lbk \frac{4xy \fz - 4 (x^{2} + y^{2})
\fz^{2}}{1-4 \fz^{2}} \rbk \right|_{\fz = 0}
\ed
in the last equality, which permits us to reach, after some manipulations, the intermediate result
\be
\lb{a4}
1 = \mathrm{N}_{n}^{2} \, \left. \frac{\upartial^{n}}{\upartial \fz^{n}} (1-4 \fz^{2})^{-1/2} \mathrm{J}(\fz) \right|_{\fz = 0} ,
\ee
with $\mathrm{J}(\fz)$ given by
\bd
\mathrm{J}(\fz) = \sum_{\{ \bet,\bet^{\prime} \} \in \mathbb{Z}} \exp \lbk - \frac{2 \pi}{N} \frac{1+4 \fz^{2}}{1-4 \fz^{2}} \lpar \bet
+ \frac{\bet^{\prime} N}{2} \rpar^{2} - \frac{\pi N}{2} \frac{1+4 \fz^{2}}{1-4 \fz^{2}} {\bet^{\prime}}^{2} + \frac{2 \pi}{N} 
\frac{4 \fz}{1 - 4 \fz^{2}} \bet (\bet + \bet^{\prime} N) \rbk .
\ed
Note that the sum over $\bet^{\prime}$ can be separated out in two contributions coming from the even $(e)$ and odd $(o)$ integers, and
consequently this procedure implies that $\mathrm{J}(\fz) = \mathrm{J}_{e}(\fz) + \mathrm{J}_{o}(\fz)$. Now, let us go one step further
in order to determine each term separately.

For example, the even term $\mathrm{J}_{e}(\fz)$ can be dealt with by shifting the sum over $\bet$ by $- \bet^{\prime} N$. This fact
produces a decoupling between the discrete indices $\bet$ and $\bet^{\prime}$, namely
\bd
\mathrm{J}_{e}(\fz) = \sum_{\bet \in \mathbb{Z}} \exp \lpar - \frac{2 \pi}{N} \frac{1-2 \fz}{1+2 \fz} \bet^{2} \rpar \sum_{\bet^{\prime}
\in \mathbb{Z}} \exp \lpar - 2 \pi N \frac{1+2 \fz}{1-2 \fz} {\bet^{\prime}}^{2} \rpar ,
\ed
which allows us to identify each sum with a particular Jacobi theta function as follows \cite{r25}:
\be
\lb{a5}
\mathrm{J}_{e}(\fz) = \lbk - \nc f(\fz) \rbk^{1/2} \vartheta_{3} [ 0 | 4 f(\fz) ] \vartheta_{3} [ 0 | f(\fz) ] ,
\ee
where $f(\fz) = \nc \fa (1-2\fz)(1+2\fz)^{-1}$ and $\fa = (2N)^{-1}$. The odd term $\mathrm{J}_{o}(\fz)$ can also be determined by means
of a similar mathematical procedure, yielding as result the analytical expression
\be
\lb{a6}
\mathrm{J}_{o}(\fz) = \lbk - \nc f(\fz) \rbk^{1/2} \vartheta_{2} [ 0 | 4 f(\fz) ] \vartheta_{4} [ 0 | f(\fz) ] .
\ee
Therefore, with the help of (\ref{a4}), the sum of these contributions leads us to obtain a closed form for the normalization constant
$\mathrm{N}_{n}$.

\subsection{Generalized function}

Initially, let us substitute the analytical expressions of the coefficients $\mathfrak{F}_{\sigma,n}$ and $\mathfrak{F}_{\sigma -
\xi,n}^{\ast}$ into Eq. (\ref{a1}). The sum over $\sigma$ yields the result $N \delta_{\alf,\bet + \eta}^{[\ns]}$, which implies that
$\alf$ will assume the values $\bet + \eta + \bet^{\prime} N$ for $\bet^{\prime} \in \mathbb{Z}$. Thus, with the help of the generating
function for the product of two Hermite polynomials and after some algebraic manipulations, the generalized function $\mathscr{K}_{n}
(\eta,\xi)$ can be written as $\mathrm{N}_{n}^{2} \mathscr{M}_{n}(\eta,\xi)$, where $\mathscr{M}_{n}(\eta,\xi)$ is defined through the
equation
\be
\lb{a7}
\mathscr{M}_{n}(\eta,\xi) \coloneq \sqrt{\fa} \, \left. \frac{\upartial^{n}}{\upartial \fz^{n}} ( 1 + 2 \fz )^{-1} \mathrm{M}
(\eta,\xi;\fz) \right|_{\fz = 0}
\ee
and $\mathrm{M}(\eta,\xi;\fz)$ denotes an auxiliary function given by
\brr
\lb{a8}
& & \mathrm{M}(\eta,\xi;\fz) = \lbk - \nc f(\fz) \rbk^{-1/2} \exp \lpar - \frac{\nc \pi}{N} \eta \xi - \frac{\pi}{N} \frac{1+4 \fz^{2}}
{1-4 \fz^{2}} \eta^{2} \rpar \sum_{\bet \in \mathbb{Z}} \exp \lpar \frac{2 \pi}{N} \frac{1+4 \fz^{2}}{1-4 \fz^{2}} \eta \bet +
\frac{2 \pi \nc}{N} \xi \bet \rpar \nn \\
& & \qquad \times \sum_{\bet^{\prime} \in \mathbb{Z}} \exp \lbk - \frac{2 \pi}{N} \frac{1+4 \fz^{2}}{1-4 \fz^{2}} \lpar \bet +
\frac{\bet^{\prime} N}{2} \rpar^{2} - \frac{\pi N}{2} \frac{1+4 \fz^{2}}{1-4 \fz^{2}} {\bet^{\prime}}^{2} + \frac{2 \pi}{N} 
\frac{4\fz}{1-4\fz^{2}} (\bet - \eta) (\bet + \bet^{\prime} N) \rbk .
\err
The next step consists basically in evaluating the sums over $\bet$ and $\bet^{\prime}$ by means of the mathematical procedure adopted
for $\mathrm{J}(\fz)$ -- \ie, the auxiliary function will be decomposed in two parts coming from the even and odd contributions
associated with the discrete index $\bet^{\prime}$.

It is worth mentioning that the technical details involved in the determination of $\mathrm{M}(\eta,\xi;\fz) = \mathrm{M}_{e}(\eta,\xi;
\fz) + \mathrm{M}_{o}(\eta,\xi;\fz)$ will be suppressed at this moment without significant consequences for our initial purpose, being
exhibited only the most important results. In this sense, the even contribution
\brr
\mathrm{M}_{e}(\eta,\xi;\fz) &=& \lbk - \nc f(\fz) \rbk^{-1/2} \exp \lpar - \frac{\nc \pi}{N} \eta \xi - \frac{\pi}{N} \frac{1+4\fz^{2}}
{1-4 \fz^{2}} \eta^{2} \rpar \sum_{\bet \in \mathbb{Z}} \exp \lbk - \frac{2 \pi}{N} \frac{1-2 \fz}{1+2 \fz} \bet^{2} + \frac{2 \pi \nc}
{N} \lpar \xi - \nc \frac{1-2 \fz}{1+2 \fz} \eta \rpar \bet \rbk \nn \\
& & \times \sum_{\bet^{\prime} \in \mathbb{Z}} \exp \lpar - 2 \pi N \frac{1+2 \fz}{1-2 \fz} {\bet^{\prime}}^{2} - 2 \pi \frac{1+2 \fz}
{1-2 \fz} \eta \bet^{\prime} \rpar \nn 
\err
assumes the simplified form
\bd
\mathrm{M}_{e}(\eta,\xi;\fz) = \half \lbr \vartheta_{3} [ \fa \eta | f(\fz) ] \vartheta_{3} [ \fa \xi | f(\fz) ] + \vartheta_{3} 
[ \fa \eta | f(\fz) ] \vartheta_{4} [ \fa \xi | f(\fz) ] \mathrm{e}^{\nc \pi \eta} \rbr ,
\ed
while the odd contribution
\brr
& & \mathrm{M}_{o}(\eta,\xi;\fz) = \lbk - \nc f(\fz) \rbk^{-1/2} \exp \lpar - \frac{\nc \pi}{N} \eta \xi - \frac{\pi}{N}
\frac{1+4 \fz^{2}}{1-4 \fz^{2}} \eta^{2} + \pi \frac{1-2\fz}{1+2\fz} \eta \rpar \nn \\
& & \qquad \times \sum_{\bet \in \mathbb{Z}} \exp \lbk - \frac{2 \pi}{N} \frac{1-2 \fz}{1+2 \fz} \lpar \bet + \frac{N}{2}
\rpar^{2} + \frac{2 \pi \nc}{N} \lpar \xi - \nc \frac{1-2 \fz}{1+2 \fz} \eta \rpar \bet \rbk \nn \\
& & \qquad \times \sum_{\bet^{\prime} \in \mathbb{Z}} \exp \lbk - 2 \pi N \frac{1+2 \fz}{1-2 \fz} \lpar \bet^{\prime} +
\frac{1}{2} \rpar^{2} - 2 \pi \frac{1+2 \fz}{1-2 \fz} \eta \lpar \bet^{\prime} + \frac{1}{2} \rpar \rbk \nn 
\err
can be represented through the expression
\bd
\mathrm{M}_{o}(\eta,\xi;\fz) = \half \lbr \vartheta_{4} [\fa \eta | f(\fz)] \vartheta_{3} [\fa \xi | f(\fz)] \mathrm{e}^{\nc \pi \xi} +
\vartheta_{4} [ \fa \eta | f(\fz) ] \vartheta_{4} [ \fa \xi | f(\fz) ] \mathrm{e}^{\nc \pi (\eta + \xi+N)} \rbr .
\ed
Note that some fundamental properties of the Jacobi theta functions were extensively used in the derivation process of
$\mathrm{M}_{e(o)}(\eta,\xi;\fz)$, such properties being promptly listed and discussed in Ref. \cite{r25}. In addition, the sum of
these contributions leads us not only to determine the function $\mathscr{M}_{n}(\eta,\xi)$ through the mathematical operation defined
in Eq. (\ref{a7}), but also to show that the normalization constant $\mathrm{N}_{n}$ is connected with $\mathscr{M}_{n}(0,0)$ by means
of the equality $\mathrm{N}_{n}^{-2} = \mathscr{M}_{n}(0,0)$. Consequently, the generalized function can be written in this context as
$\mathscr{K}_{n}(\eta,\xi) = \mathscr{M}_{n}(\eta,\xi) / \mathscr{M}_{n}(0,0)$, which allows us to obtain directly the diagonal matrix
element $\lgg n| {\bf S}(\eta,\xi) |n \rg$.

\section{The squeezing generator for finite-dimensional Hilbert spaces}

We initiate this appendix establishing some particular results for the squeezing generator $\sop(r)$ in infinite-dimensional Hilbert
spaces which constitute the continuous counterparts of the discrete cases to be presented in this section. For instance, the normalized
wavefunction for the squeezed coherent states can be obtained through a simple mathematical procedure sketched out by the relation
\be
\lb{b1}
\lgg x | \bar{q},\bar{p};r \rg = \int d x^{\prime} \lgg x | \sop(r) | x^{\prime} \rg \lgg x^{\prime} | \bar{q},\bar{p} \rg ,
\ee
where
\be
\lb{b2}
\lgg x^{\prime} | \bar{q},\bar{p} \rg = \pi^{-1/4} \exp \lbk - \frac{1}{2} ( x^{\prime} - \bar{q} )^{2} + \nc \bar{p} \lpar x^{\prime}
- \frac{\bar{q}}{2} \rpar \rbk
\ee
represents the wave function for the continuous coherent states in the coordinate representation $\{ | x^{\prime} \rg \}$, and
\be
\lb{b3}
\lgg x | \sop(r) | x^{\prime} \rg = \lam^{1/4} \exp \lbk - \frac{1}{4} ( x^{\prime} - \lam^{1/2} x )^{2} \rbk \delta ( x^{\prime} -
\lam^{1/2} x )
\ee
is the nondiagonal matrix element associated with the squeezing operator $\sop(r)$ when $\lam \coloneq \exp (-2r)$. The adequate
manipulation of these results allows us to show that (\ref{b1}) can then be promptly derived by means of the identity $\lgg x | \bar{q},
\bar{p}; \lam \rg = \lam^{1/4} \lgg \lam^{1/2} x | \bar{q},\bar{p} \rg$. In fact, this simple operation leads us to obtain a generalized
set of wave functions belonging to a wide class of squeezed states of the electromagnetic field characterized by \cite{r4} 
\bd
\lbr | \Psi;r \rg \coloneq \sop(r) | \Psi \rg : \sop(r) = \exp [ - \nc (r/2) \{ {\bf Q},{\bf P} \} ], \; r \in (-\infty,\infty) \rbr ,
\ed
namely $\lgg x | \Psi ; \lam \rg = \lam^{1/4} \lgg \lam^{1/2} x | \Psi \rg$.

Now, let us apply this result to a particular set of nonclassical states described by the number states $\{ | n \rg \}_{n \in
\mathbb{N}}$. Thus, the normalized wavefunction associated with the squeezed number states assumes the analytical form \cite{ap2}
\be
\lb{b4}
\lgg x | n;\lam \rg = \lbk \frac{\lam}{\pi ( 2^{n} n! )^{2}} \rbk^{1/4} \exp \lpar - \half \lam x^{2} \rpar H_{n}(\lam^{1/2} x) ,
\ee
whose complementary wavefunction in the momentum space representation,
\be
\lb{b5}
\lgg p_{x} | n;\lam \rg = \frac{(-\nc)^{n}}{\lbk \pi \lam ( 2^{n} n! )^{2} \rbk^{1/4}} \exp \lpar - \half \lam^{-1} p_{x}^{2} \rpar
H_{n} (\lam^{-1/2} p_{x}) ,
\ee
permits us to establish a well-known important link between both the coordinate and momentum representations via the Fourier transform,
\be
\lb{b6}
\lgg x | n;\lam \rg = \int \frac{dp_{x}}{\sqrt{2 \pi}} \exp ( \nc x p_{x} ) \lgg p_{x} | n;\lam \rg .
\ee
Basically, these results will be our guide-lines in the construction process of the discrete counterpart for the squeezing generator in
finite-dimensional Hilbert spaces.

In appendix A we have shown that $\{ \mathfrak{F}_{\kappa,n} \}$ are the coefficients that lead us to connect the finite number states
$\{ | n \rg \}_{n = 0, \ldots, N-1}$ and the discrete coordinate-like representation $\{ | u_{\kappa} \rg \}_{\kappa = - \ell,\ldots,
\ell}$ through the expansion
\be
\lb{b7}
| n \rg = \sum_{\kappa = - \ell}^{\ell} \mathfrak{F}_{\kappa,n} | u_{\kappa} \rg .
\ee
Following, let us define the squeezed number states in the context of finite-dimensional Hilbert spaces by means of the mathematical
statement
\be
\lb{b8}
| n;\fs \rg \coloneq \sqq(\fs) | n \rg = \sum_{\kappa = - \ell}^{\ell} \mathfrak{F}_{\kappa,n}(\fs) | u_{\kappa} \rg ,
\ee
where $\{ \mathfrak{F}_{\kappa,n}(\fs) \}$ represent the generalized coefficients expressed as
\brr
\lb{b9}
\mathfrak{F}_{\kappa,n}(\fs) &=& \mathrm{N}_{n}(\fs) \frac{(-\nc)^{n}}{\sqrt{N}} \sum_{\bet \in \mathbb{Z}} \exp \lpar - \frac{\pi}{N}
\fs^{-2} \bet^{2} + \frac{2 \pi \nc}{N} \bet \kappa \rpar H_{n} \lpar \sqrt{\frac{2 \pi}{N}} \, \fs^{-1} \bet \rpar \nn \\
&=& \frac{\mathrm{N}_{n}(\fs)}{\sqrt{( \pi \fa \fs^{2})^{n} N}} \left. \frac{\upartial^{n}}{\upartial \fz^{n}} \exp (\pi \fa \fs^{2}
\fz^{2}) \, \vartheta_{3} \lbk 2 \fa ( \kappa - \fz ) | 2 \nc \fa \fs^{-2} \rbk \right|_{\fz = 0} ,
\err
$\mathrm{N}_{n}(\fs)$ being the normalization constant given by
\bd
\mathrm{N}_{n}^{-2}(\fs) = \sqrt{\fa \fs^{2}} \, \frac{\upartial^{n}}{\upartial \fz^{n}} \, (1 + 2 \fz)^{-1} \lbr \vartheta_{3} 
[ 0 | 4 \fs^{-2} f(\fz) ] \vartheta_{3} [ 0 | \fs^{2} f(\fz) ] + \vartheta_{2} [ 0 | 4 \fs^{-2} f(\fz) ] \vartheta_{4} [ 0 | \fs^{2}
f(\fz) ] \rbr \Biggr|_{\fz = 0} . 
\ed
Note that the squeezing parameter $\fs$ plays, in principle, the same functional role reserved to the continuous counterpart $\lam$ --
compare Eqs. (\ref{b4}) and (\ref{b9}) -- i.e., both the squeezing parameters modify the widths of their respective wave packets and
this fact allows us to introduce satisfactorily the squeezing effects within the context of finite-dimensional Hilbert spaces. In
addition, the wavepacket for the discrete squeezed-vacuum state (\ref{e8}) can be considered as a particular component of the
generalized coefficients since it corresponds to the specific case $\mathfrak{F}_{\kappa,0}(\fs)$. Consequently, the complete set of
components characterized by $\{ \mathfrak{F}_{\kappa,n}(\fs) \}_{n=0,\ldots,N-1}$ will be sufficient in this approach to determine
precisely the squeezing operator $\sqq(\fs)$ and its inherent mathematical properties.

The action of $\sqq(\fs)$ on the finite number states not only defines the squeezed number states (\ref{b8}) but also leads us to
establish a formal expression for the squeezing generator as follows:
\be
\lb{b10}
\sqq(\fs) \coloneq \sum_{n=0}^{N-1} | n;\fs \rg \lgg n | .
\ee
In what concerns the operator $\sqq(\fs)$, it is worth noticing that its representation in the set of Mehta's states \cite{r26} carries
a peculiar behaviour inherent to those states. Indeed the study of this particular representation deserves a detalied investigation
since the peculiarities mentioned before come from nonorthogonality of the Mehta's set -- for technical details on theoretical and
numerical investigations of this specific topic, see Ref. \cite{r27}. Finally, let us mention briefly that the proof on the unitarity 
of the discrete squeezing operator will be published elsewhere \cite{ap3}.
\newpage


\end{document}